
%
%
%
%

\def\pmb#1{\setbox0=\hbox{$#1$}%
\kern-.025em\copy0\kern-\wd0
\kern.05em\copy0\kern-\wd0
\kern-.025em\raise.0433em\box0 }
\def\tr{\,{\rm tr}\,}
\def\lapp{\hbox{$ {     \lower.40ex\hbox{$<$}
                   \atop \raise.20ex\hbox{$\sim$}
                   }     $}  }
\def\rapp{\hbox{$ {     \lower.40ex\hbox{$>$}
                   \atop \raise.20ex\hbox{$\sim$}
                   }     $}  }
\def\diag{\,{\rm diag}\,}
\magnification=1200
\hoffset=-.1in
\voffset=-.2in

\vsize=7.5in
\hsize=5.6in
\tolerance 10000

\baselineskip 12pt plus 1pt minus 1pt
\pageno=0
\baselineskip 12pt plus 1pt minus 1pt
\centerline{\bf MINKOWSKI SPACE NON-ABELIAN CLASSICAL
SOLUTIONS}
\smallskip
\centerline{{\bf WITH NON-INTEGER WINDING NUMBER CHANGE}
\footnote{*}{This work is supported in part by funds
provided by the U. S. Department of Energy (D.O.E.) under
contract
\#DE-AC02-76ER03069, and in part by funds provided by the
TNRLC under
grant \#RGFY92C6.}}
\vskip 24pt
\centerline{Edward Farhi, Valentin V. Khoze and Robert
Singleton, Jr.
\footnote{**}{SSC Fellow}}
\vskip 12pt
\centerline{\it Center for Theoretical Physics}
\centerline{\it Laboratory for Nuclear Science}
\centerline{\it and Department of Physics}
\centerline{\it Massachusetts Institute of Technology}
\centerline{\it Cambridge, Massachusetts\ \ 02139\ \ \
U.S.A.}
\vskip 1.5in
\centerline{Submitted to: {\it Physical Review D}}
\bigskip\bigskip
\centerline{\bf ABSTRACT }
Working in a spherically symmetric ansatz in Minkowski space
we discover new solutions to the classical equations of motion
of pure $SU(2)$ gauge theory.
These solutions represent spherical shells of energy which
at early times move inward near the speed of light, excite the
region of space around the origin at intermediate times and
move outward at late times. The solutions change the winding
number in bounded regions centered at the origin by non-integer
amounts. They also produce non-integer topological charge in
these regions. We show that the previously discovered solutions
of L\"uscher and Schechter also have these properties.
\vfill

\centerline{ Typeset in $\TeX$ by Roger L. Gilson}
\vskip -12pt
\noindent CTP\#2138  \hfill December 1992
\eject
\baselineskip 24pt plus 2pt minus 2pt

\noindent{\bf I.\quad INTRODUCTION}
\medskip
In non-Abelian gauge theories coupled to fermions, gauge field
configurations with non-trivial topological charge can cause explicit
violation of conservation laws as a result of anomalous Ward
identities.$^1$
This leads to the violation of chirality in QCD and
non-conservation of baryon and lepton number in the
electroweak theory.$^2$ The traditional approach
to understanding these effects uses semiclassical barrier penetration
where the tunneling solutions are Euclidean instantons.$^3$
In the electroweak theory the height of the barrier is  of order
$M_{\rm w}/\alpha_{\rm \scriptscriptstyle w} \sim 10 {\rm TeV}$ --
the energy of the sphaleron$^4$ --
and at energies much below this, fermion number violation
is exponentially suppressed.

More recently it has been suggested that fermion number violating processes
may become unsuppressed if high temperatures$^5$ or high energies$^6$
are involved. The idea is that instead of tunneling through the
barrier the field configuration can pass over it.

At temperatures above the barrier height it is generally assumed that
field configurations which pass over the barrier can be found in the
thermal ensemble.
The rate estimates proceed through thermodynamic arguments which
are fairly general and make reference to little more than the temperature,
barrier height and coupling constant.

At high energies the existing techniques are based mostly on
Euclidean methods which deal with contributions of instanton-like
configurations to the functional integral (for recent reviews see
Ref.~[7]). These essentially semiclassical methods have intrinsic
problems at energies of order the barrier height, exactly where anomalous
fermion number production may become unsuppressed.

We are pursuing a complementary approach$^8$ based entirely
in Minkowski space
which corresponds to passage over the barrier. Our approach may be
separated into two main parts: the creation and evolution of finite energy
classical gauge fields by particle collisions and the dynamics of fermion
number (or chirality) violation in the presence of such background gauge
sources.
In this paper we study the classical evolution of certain gauge field
configurations in Minkowski space. The problem of fermion production is
addressed in a companion paper.$^9$

In the Euclidean approach what is typically done is to only consider
finite action contributions to the functional integral.
These configurations fall into homotopy classes distinguished by an integer,
$Q$, called the topological charge, given by

$$ Q = {g^2 \over 16 \pi^2}
\ \int d^4x~ {1\over 2}\epsilon^{\mu\nu\alpha\beta}{\rm tr}
  (F_{\mu \nu} F_{\alpha \beta}) \ .
\eqno(1.1)$$
What we discuss here are Minkowski space classical solutions that
produce arbitrary non-integer topological charge and winding number
change.

Consider a spherical region of space with radius $R$ containing
vanishing gauge fields at some large negative time, $-T_0$.
Imagine sending in a localized spherical wave-front of energy
$E$. At time $t \sim -R$ the front first reaches the region. Energy flows in,
bounces back at time $t\sim 0$, and then leaves the region at time
$t\sim R$. At a large positive time, $+T_0$, the region of space interior to
$R$ is once again in a pure gauge, but now the vector potential $A_\mu$
need not vanish. The question we address is, what is the topological structure
of the pure gauge configuration left behind by the wave-front?
We will construct solutions to the Yang-Mills equations which describe
such localized wave-fronts and study in detail the winding
number of the pure
gauge configurations that the fronts leave behind. In addition
we compute the topological charge created inside the region.
We will also study the topological properties of the previously
discovered solutions of L\"uscher$^{10}$  and Schechter,$^{11}$
which we also show describe localized incoming and outgoing spherical waves.

Classical vacuum configurations are typically classified$^{12}$
by the winding number of the associated gauge function.
In the $A_0=0$ gauge the classical vacuum configurations
are of the form
$A_i =i/g \ U \partial_i U^\dagger$
where $U({\bf x})$, the gauge function, is a map from the $R^3$ into
the gauge group. The winding number is
$$
  \nu\lbrack U \rbrack=
  {1 \over 24 \pi^2} \int d^3{\bf x} \ \epsilon^{ijk} \
  {\rm tr} (U^\dagger \partial_i U \ U^\dagger \partial_j U \
  U^\dagger \partial_k U ) \
.\eqno(1.2)
$$
If we assume that $U({\bf x}) \to 1$ as $\mid {\bf x}\mid \to \infty$
then $U({\bf x})$ is actually a map from $S^3$ into the gauge group.
These maps can be classified by an integer which
labels the homotopy class of the map. Formula (1.2) is an explicit
expression for this integer.
The standard picture$^{12}$ of the gauge theory vacuum is based upon this
classification. The $\theta$-vacua of QCD are linear superpositions of
states associated with classical vacuum configurations which have
integer winding number. However, if $U({\bf x})$ is not required to
approach a constant at spatial infinity, such an elegant vacuum
classification is not available.
One typically argues, however, that by causality  the values of fields
at spatial infinity are unimportant, so imposing a convenient boundary
condition does not entail a loss of generality. Also, instantons can be
viewed as changing winding number by an integer amount. If only
instanton-like configurations are important, then restricting to vacua with
integer winding number can be justified after the fact.

However, if we return to the description of the spherical region of space
with radius $R$ that is excited by a wave-front and then relaxes back to
pure gauge, we find that $U$ at $R$ does not equal a direction-independent
constant. Accordingly, the local winding number at late times, computed by
doing the integral in (1.2) only out to $R$, does not equal an integer even if
$R$ is very large.
That is to say, if our known universe is the region in question,
``aliens'' from the outside could send in a wave which could leave our
universe in a different classical vacuum, one which did not belong to the
previous topological classification.

If the wave which visits the region changes its winding number by a
non-integer amount, and creates non-integer topological charge, we can
ask what the implications are for anomalous fermion production
in the region. In a companion paper$^9$ we study fermion number
production in a (1+1)-dimensional theory with an arbitrary background
gauge field. We arrange for the background field
to change its winding number in local regions and we allow for these local
changes to be non-integer. We find that fermions are produced only in these
local regions. Furthermore we discover that the the number of fermions produced
in a local region, in a quantum average sense, is equal to the change in
winding
number of the gauge field in this region.
That is, if in a local region the change in winding number is $f$, and the
background field is used time and time again to quantum-mechanically produce
fermions, then in the local region with each observation you will find
integer net fermion production, but the quantum average will be $f$.
This suggests that if
we restrict our attention to the region of radius $R$, and a spherical
non-Abelian wave comes in and leaves, there will be fractional fermion number
production in the sense described above. Here we are discussing net fermion
number production in the region, that is the difference between the number
of fermions coming into the region through the two-sphere of radius $R$
and the number leaving.

The outline of the paper is as follows. In Section II we explain the
spherical ansatz$^{13}$ which we use to simplify the Yang-Mills
equations and we introduce a further ansatz which allows us
to find new solutions. In Section III we study in our language
the solutions previously found by L\"uscher$^{10}$ and Schechter.$^{11}$
We show that these solutions have many of the qualitative features
of the solutions of Section II.
In both cases the solutions have certain transformation properties
under $SO(2,1)$ which allows us to generate new solutions. The action
of symmetry transformations on the solutions is discussed in Section IV.
In Section V we investigate the topological properties of both classes of
solutions. Finally in Section VI we discuss further questions which
arise from this work.
\goodbreak
\bigskip
\noindent{\bf II.\quad THE SPHERICAL ANSATZ AND NEW YANG-MILLS SOLUTIONS}
\medskip
\nobreak
In this section we consider the spherical ansatz$^{13}$
for $SU(2)$ gauge theory. We follow the notation of
Ref.~[14].  We find new classical solutions in (3+1)-dimensional
Minkowski space and discuss their properties.

The action for pure $SU(2)$ gauge theory is
$$
  S =  -{1\over 2} \int d^4x\tr \left( F_{\mu\nu}F^{\mu\nu}\right)\ \
,\eqno(2.1)$$
where $F_{\mu\nu}= F^a_{\mu\nu} \left(\sigma^a/2\right) = \partial_\mu
A_\nu - \partial_\nu A_\mu - i g \left[ A_\mu, A_\nu\right]$ is the field
strength and $A_\mu= A^a_\mu\left(\sigma^a/2\right)$.  Throughout we use the
space-like convention for the metric tensor,
$  \eta_{\mu\nu} = \diag (-1,+1,+1,+1)$,
The spherical  ansatz is given in terms of the four functions
$a_0,a_1,\alpha,\beta$ by
$$\eqalign{
  A_0 ({\bf x},t) &= {1\over 2g} ~ a_0 (r,t) {\pmb\sigma}\cdot
  \hat{\bf x}\ \ ,\cr
  A_i({\bf x},t) &= {1\over 2g}\  \left( a_1 (r,t) e^3_i +
  {\alpha(r,t)\over
  r} e^1_i + {1+\beta(r,t)\over r} e^2_i\right)\ \ ,\cr}\eqno(2.2)$$
where the matrix-valued functions $\{e^k_i\}$ are defined as
$$\eqalign{
  e^1_i &= \sigma_i - {\pmb\sigma} \cdot \hat{\bf x}\hat x_i\ \ ,\cr
  e^2_i &=  i \left[{\pmb\sigma} \cdot \hat{ \bf x}\sigma_i - \hat
  x_i\right] = \epsilon_{ijk} \hat x_j \sigma_k \ \ ,\cr
  e^3_i &=  {\pmb\sigma} \cdot\hat{\bf x} \hat x_i\ \ ,\cr}\eqno(2.3)$$
and where $\hat{\bf x}$ is a unit three-vector in the radial direction.

The action (2.1) in the spherical  ansatz takes the form
$$  S = {4\pi\over g^2} \int dt \int^\infty_0 dr \left( -{1\over 4} r^2
  f_{\mu\nu} f^{\mu\nu} - \left( D_\mu \chi\right)^* D^\mu \chi - {1\over
  2r^2}\left(|\chi|^2 - 1 \right)^2\right)\ \ .\eqno(2.4)$$
The $(1+1)$-dimensional field strength is defined as $f_{\mu\nu} =
\partial_\mu a_\nu - \partial_\nu a_\mu$, where $\mu,\nu=t,r$, and
indices are raised and lowered with $\tilde\eta_{\mu\nu}={\rm diag}(-1,+1)$.
A scalar field  $\chi = \alpha+i\beta$, with covariant derivative $D_\mu
\chi = \left( \partial_\mu - i a_\mu\right)\chi$, has also been introduced.

The  ansatz (2.2) preserves a residual $U(1)$ subgroup of the $SU(2)$
 gauge group consisting of the transformations,
$$  U({\bf x},t) = \exp \left[i\Omega(r,t) {{\pmb \sigma}\cdot\hat{\bf
  x}\over 2} \right]\  .\eqno(2.5)$$
These induce the gauge transformations
$$  a_\mu \to a_\mu + \partial_\mu \Omega\ \ ,\qquad \chi \to \exp (i\Omega)
  \chi \ \ ,\eqno(2.6)$$
which are seen to leave (2.4) invariant.

Vacuum configurations in $(3+1)$-dimensions in the $A_0 = 0$ gauge are given
by the pure gauges,
$$
  A_i^{(\rm vac)} =i/g \ U \partial_i U^\dagger \ \ ,\eqno(2.7)$$
where $U$ is a time-independent function from three-space into $SU(2)$.
If we wish to maintain spherical symmetry then
$$
  U(r) = \exp\left[i\Omega(r)~{{\pmb\sigma}\cdot\hat{\bf x} \over
  2}\right]  \ \ .\eqno(2.8)$$
\def\vac{{\rm vac}}
In terms of the reduced theory we have from (2.7) and (2.2) that
$$
  a^\vac_0 =0 \  \ ;\ \ a^\vac_1 = {d\Omega \over dr} \  \ ;\ \
  \chi^\vac=-ie^{i\Omega}\ \ .\eqno(2.9)$$
This form for the vacuum configurations in the (1+1)-dimensional
theory could easily be obtained from the action (2.4).

In (3+1) dimensions the function (2.8) is well-defined at $r=0$ only
if $\Omega(0)=2\pi n$ with $n$ an integer. If the further restriction
is imposed that $U \to 1$ at spatial infinity, then we have
$\Omega(\infty)=2\pi m$ with $m$ an integer. The winding number of this
configuration, as computed by (1.2), is $m-n$. If we impose  no boundary
condition at spatial infinity, then $U$ would not have an integer
winding number. From the point of view of the reduced theory, the vacuum
configuration is given by (2.9) and we see no reason, at this point, to
impose restrictions on $\Omega$ at zero or infinity. We will see,
however, that a boundary condition at $r=0$ (but not at $r=\infty$)
will be dynamically
imposed on all finite energy solutions to the equations of motion.

The $(1+1)$-dimensional equations of motion for the reduced theory (2.4)
are given by
$$  - \partial^\mu \left(r^2 f_{\mu\nu} \right) = i \left[ \left( D_\nu
  \chi\right)^*\chi - \chi^* D_\nu \chi\right]\ ,\eqno(2.10\hbox{a})$$
$$  \left( - D^2 + {1\over r^2} \left( |\chi|^2-1\right)\right) \chi = 0
\ . \eqno(2.10\hbox{b})$$
Let us express the complex scalar field $\chi$ in polar form,
$$  \chi(r,t) = -i \rho(r,t) \exp \left[ i \varphi(r,t)\right]\ \
,\eqno(2.11)$$
where $\rho$ and $\varphi$ are real scalar fields. We must be
careful in (2.11) when $\rho$ vanishes, since the
angle $\varphi$ can change discontinuously by $\pi$ if we require
that $\rho \ge 0$. We will choose the sign of $\rho$ to ensure the
continuity of $\varphi$ as $\chi$ varies smoothly through zero.
In terms of $\rho$, $\varphi$ and $a_\mu$, the four equations contained
in (2.10) read
\vfil\eject
$$
  \partial^\mu\left( r^2 f_{\mu\nu}\right) + 2\rho^2 \left(
  \partial_\nu\varphi  - a_\nu\right) = 0\ \ , \eqno(2.12\hbox{a})
$$
$$
  \partial^\mu \partial_\mu \rho - \rho\left( \partial^\mu \varphi-
  a^\mu  \right)\left( \partial_\mu \varphi -a_\mu  \right)
 - {1\over r^2} \rho \left(\rho^2-1\right) = 0 \ \ ,
\eqno(2.12\hbox{b})
$$
and
$$
  \partial^\mu \big[\rho^2(\partial_\mu \varphi - a_\mu)\big]=0 \ .
\eqno(2.12\hbox{c})
$$
The last equation follows from (2.12a) so we see there are only three
independent equations, not four, as we expect because of gauge invariance.
Since in (1+1) dimensions $f_{\mu\nu}$ must be proportional to
$\epsilon_{\mu\nu}$, it is convenient to define a new field $\psi$ by
$$  r^2 f_{\mu\nu} = - 2 \epsilon_{\mu\nu}\psi\ \ ,\eqno(2.13)$$
where $\epsilon_{\mu\nu}$ is the antisymmetric symbol with $\epsilon_{01}=+1$.
Contracting (2.13) with $\epsilon^{\mu\nu}$ gives the relation
$$  \epsilon^{\mu\nu} \partial_\mu a_\nu = {2 \over r^2} \psi\ \
,\eqno(2.14)$$
which we use below.
By using (2.13) in (2.12a) and contracting with
$\epsilon^{\nu\alpha}$ we find
$$  \partial^\alpha \psi = - \epsilon^{\alpha\nu}\rho^2(\partial_\nu
  \varphi - a_\nu)\ \ .\eqno(2.15)$$
Notice that (2.15) implies (2.12c). From this last equation it
follows that
$$  \partial_\alpha\bigg({\partial^\alpha \psi \over \rho^2}\bigg) =
  \epsilon^{\alpha\nu}\partial_\alpha a_\nu = {2\over r^2} \psi \ ,
\eqno(2.16)$$
where we have used (2.14) in passing to the last equality. This gives an
equation solely in terms of the fields $\rho$ and $\psi$. We may also use
(2.15) to express the second term in (2.12b) in terms of only $\rho$
and $\psi$.  We then have the alternate classical equations of motion
\vfil\eject
$$  -\partial_t^2\rho +\partial_r^2\rho
  - {1\over \rho^3}~(\partial_t\psi)^2
  + {1\over \rho^3}~(\partial_r\psi)^2
  - {1\over r^2}\rho(\rho^2-1) = 0 \ ,
\eqno(2.17\hbox{a})
$$
$$
  -\partial_t \big({\partial_t \psi \over \rho^2} \big)
  +\partial_r \big({\partial_r \psi \over \rho^2} \big)
  - {2\psi \over r^2} = 0 \ .
\eqno(2.17\hbox{b})
$$
Equations (2.17)  are equivalent to the original Eqs.
(2.10), but for our purposes they will be more convenient.
Note that $\rho$ and $\psi$ are gauge invariant fields, and we have only two
equations.

Using the equations of motion, the energy associated with the action
(2.4) can be written in terms of $\rho$ and $\psi$ as
$$
  \eqalign{E &= {8\pi \over g^2}~
  \int^\infty_0 dr\biggl[ {1\over 2} \left(\partial_t
  \rho\right)^2 + {1\over 2}\left( \partial_r \rho\right)^2
  +{1\over 2\rho^2} \left( \partial_t\psi\right)^2 \cr &\qquad+ {1\over
  2\rho^2} \left(\partial_r\psi\right)^2 + {\psi^2\over r^2} + {\left(
  \rho^2-1\right)^2\over4r^2}~\biggr]\ \ .\cr}
\eqno(2.18)$$
We are interested in finding finite energy solutions to
(2.17). Thus the form of $\rho$ and $\psi$ when $r\sim 0$
must be
$$\eqalignno{
  \rho &=  1 + {\cal O}(r^{1/2~+~\lambda}) \ ,
                                            &(2.19\hbox{a}) \cr
  \psi &= {\cal O}(r^{1/2~+~\gamma}) \ ,
                                      &(2.19\hbox{b})
  }
$$
where $\lambda,\gamma > 0$. We have used the symmetry $\rho\to
-\rho$ under which (2.17) is invariant to fix $\rho$ to be at
$+1$ and not $-1$ as $r$ goes to zero.

Witten$^{13}$ observed that (2.4) is the action for an Abelian Higgs
model on a surface of constant curvature.  To see this in terms of $\rho$
and $\psi$ fields we can write the equations of motion (2.17) in
covariant form
\vfil\eject
$$
  {1\over \sqrt{g} }~\partial_\mu \left(\sqrt{g}~g^{\mu\nu} \partial_\nu
  \rho \right) +
  {1\over \rho^3}~
  g^{\mu\nu}\partial_\mu \psi\partial_\nu \psi - \rho(\rho^2-1) = 0 \ \ .
\eqno(2.20\hbox{a})
$$
$$
  {1\over \sqrt{g} }~\partial_\mu \left(\sqrt{g}~g^{\mu\nu}{\partial_\nu
  \psi \over \rho^2} \right)- 2 \psi = 0 \ \ ,
\eqno(2.20\hbox{b})
$$
where $g_{\mu\nu}=\tilde\eta_{\mu\nu}/r^2$ and
$g=|{\rm det} g_{\mu\nu}|$. (Recall that
the covariant derivative of a vector field  can be
written as $\nabla_\mu V^\mu=g^{-1/2}~\partial_\mu (g^{1/2} ~V^\mu)$.)
The metric $g_{\mu\nu}$
is the metric for two dimensional De Sitter space.
To see this, consider the hyperboloid $z_0^2-z_1^2-z_2^2=-1$ where the $z_i$
are parametrized in terms of $r$ and $t$ as
$$\eqalignno{
  z_0 &= {1 - r^2 +t^2 \over 2r}\ \ ,
  &(2.21\hbox{a}) \cr\noalign{\vskip 0.2cm}
  z_1 &= {t\over r}\ \ ,
  &(2,21\hbox{b}) \cr\noalign{\vskip 0.2cm}
  z_2 &= {1 + r^2 - t^2 \over 2r}\ \ .
  &(2.21\hbox{c}) \cr}
$$
It is straightforward to check that $ds^2=d z_0^2-d z_1^2-
d z_2^2=(-dt^2+dr^2)/r^2$. The coordinates $r$ and $t$
cover only half of the hyperboloid, $H^+$, for which $z_0+z_2>0$.
This is shown in Fig.~1.
The $z$-coordinates (2.21) will be useful
for studying symmetry properties of solutions. To obtain our solutions,
however, it is more convenient to work with coordinates
$w$ and $\tau$ that live on the  hyperboloid itself, and in fact
cover the whole hyperboloid:
$$\eqalignno{
  z_0 &=-\tan w \ ,          &(2.22\hbox{a})  \cr
  z_1 &=\sin \tau/ \cos w \ ,&(2.22\hbox{b})  \cr
  z_2 &=\cos \tau/ \cos w \ ,&(2.22\hbox{c})  \cr
  }
$$
where $|\tau| \le \pi$ and $|w|< \pi/2$. The coordinate $w$ is a
bounded measure of the longitudinal position along the hyperboloid
and $\tau$ measures the azimuthal angle. Notice that for fixed
$t$, as $r$ varies from zero to infinity, $w$ varies from
$-\pi/2$ to $+\pi/2$. In terms of $w$-$\tau$ variables, the
metric is $ds^2=(-d\tau^2 +dw^2)/\cos^2w$, so
(2.20) takes the form
\vfil\break
$$
  -\partial_\tau^2\rho + \partial_w^2\rho - {1\over \rho^3}
  \left(\partial_\tau \psi \right)^2 + {1\over \rho^3}\left(
  \partial_w \psi\right)^2 - {\rho \left(\rho^2-1\right) \over
  \cos^2 w} = 0 \ ,
\eqno(2.23\hbox{a})
$$
$$
  -\partial_\tau \left({\partial_\tau \psi \over \rho^2}\right) +
  \partial_w  \left({\partial_w\psi \over \rho^2}\right) -
  {2\psi \over \cos^2 w} = 0 \ .
\eqno(2.23\hbox{b})
$$

We make the  ansatz that there are finite energy solutions
to (2.23) which are independent of $\tau$. These obey the
ordinary differential equations,
$$
  {d^2\rho\over dw^2} + {1\over \rho^3}\left( {d\psi\over
  dw}\right)^2 - {\rho(\rho^2-1)\over \cos^2w} = 0\ \ ,
\eqno(2.24\hbox{a})
$$
$$
  {d\over dw} \left( {1\over \rho^2} \ {d\psi\over dw} \right)
  -{2\psi\over \cos^2w} = 0 \ \ .
\eqno(2.24\hbox{b})
$$

It is interesting to note that the previously discovered solution
of de~Alfaro, Fubini and Furlan$^{15}$ can immediately
be obtained in the following form:
$$
  \psi = 0\ \ ,\qquad \rho = -\sin w\ \ . \eqno(2.25)$$

Before we consider the full coupled system (2.24), we examine
the $\psi=0$ sector where there are solutions other than (2.25).
Equation (2.24a) with $\psi=0$ describes a ``particle'' rolling
in a ``time'' dependent potential
$V=-(\rho^2-1)^2/4\cos^2w$. We can make a change of variables from
$w$ to $\eta$, defined by $\sin w = \tanh \eta$, which removes the
``time'' dependence from the potential at the price of introducing a
``friction'' term:
$$
  \ddot \rho ~+~ \tanh\eta ~\dot \rho  ~-~ \rho(\rho^2-1) = 0\ \
.\eqno(2.26)$$
The dot denotes differentiation with respect to $\eta$, and
$\eta$ runs from $-\infty$ to $+\infty$. Now (2.26)  has a mechanical
analogue to a ``particle'' rolling in a potential $U=-{1\over 4}(\rho^2-1)^2$
with a ``frictional'' force $~-\tanh\eta ~\dot\rho$. The ``energy'' of the
mechanical system is defined by $T={1\over 2}\dot\rho^2 + U$, and for a
particle satisfying the equations of motions, $\dot T=-\tanh\eta~\dot\rho^2$.
So for negative $\eta$ the frictional force pumps energy into the system and
for positive $\eta$ it pumps energy out of the system. We are interested
in starting the particle at  $\rho(-\infty)= 1$ because of (2.19a).
The particle can only stop rolling at an extremum of the potential
$U$ so that $\rho(+\infty)=\pm 1,0$. If the particle does not stop
rolling,  $\rho$ grows without bound which is unacceptable if we
wish (2.18) to remain finite. If the energy pumped into the
system at negative $\eta$ exactly matches the dissipation at positive
$\eta$, then $\rho$ can climb up the other hump at $\rho=-1$, as does the
solution (2.25). Typically this does
not happen. If the velocity of the particle when it reaches the
bottom of the potential is too large the particle  overshoots
the other hump and $\rho$ blows up at finite $\eta$. However, if
the velocity of the particle is not too large, it gets trapped at
the bottom of the potential and dissipation drives $\rho$ to zero.
An example of a numerically generated solution to (2.26) which
goes from $\rho=1$ to $\rho=0$ is shown in Fig.~2.
We can specify a general trajectory by
choosing $\rho$ and $\dot\rho$ at $\eta=0$, giving a two parameter
family of solutions. For a
given $\rho(0)$, we must fine-tune $\dot\rho(0)$
so that $\rho(\eta)\to 1$  as $\eta\to -\infty$. This freezes out a
degree of freedom and gives a one parameter family of solutions. The future
behavior is typically to have $\rho$ dynamically driven to zero.
However with further fine-tuning we can arrange for $\rho$ to approach
$-1$ at late times, which is the solution (2.25).

At first sight making the $\tau$-independent ansatz for
(2.23) looks similar to making the assumption
that solutions to (2.17) exist which are independent of time.
It is known that such static solutions do not exist.$^{16}$
We illustrate this by considering (2.17a) with
$\psi=0$ and assuming $\partial\rho/\partial t = 0$. Making the change
of variables $s={\rm ln} r$, the equation of motion for $\rho$ becomes
$$
  \rho'' ~-~ \rho' ~-~\rho(\rho^2-1) = 0\ \ ,\eqno(2.27)$$
where the prime denotes differentiation with respect to $s$.
This represents a particle rolling in an
inverted potential $U=-{1\over 4}(\rho^2-1)^2$ with a constant frictional
force $+\rho'$. For a particle satisfying the equations of
motion, the power input is $T' = \rho'^2$. Thus energy is always pumped
into the system, and a particle that starts at $\rho=1$ will always
overshoot the other hump. Thus, even though $\tau$ and $t$ are both
time-like, making the static $\tau$ ansatz in (2.24) is in fact
quite different from making the static $t$ ansatz in (2.17).

We are now interested in finding finite energy solutions
to (2.24) with non-trivial $\psi$-dependence. Recall that
$-\pi/2 < w < \pi/2$ and that for fixed $t$ as $r\to 0$,
$w\to -\pi/2$ and as $r\to\infty$ we have $w\to\pi/2$. In
the appendix we investigate the form of finite energy solutions
near the end points of the range of $w$.
We find the following asymptotic
behavior for $w=-\pi/2 + \epsilon$ with $0<\epsilon<<1$,
$$\eqalignno{
  \rho &=  1 + {\cal O}(\epsilon^2) =  1 + {\cal O}(r^2) \ ,
  &(2.28a)\cr
  \psi &= {\cal O}(\epsilon^2)= {\cal O}(r^2) \ .
  &(2.28b)\cr}$$
Note that these conditions, imposed by dynamics,
are stronger than the finite energy conditions (2.19) \ .

We also show that for finite energy solutions in
which $\psi$ is not identically zero, $\rho$ is dynamically
driven to zero and undergoes an infinite number of oscillations as
$w \to \pi/2$. Writing $w=\pi/2 - \epsilon$ with $0<\epsilon<<1$,
the form of such solutions to leading order in $\epsilon$ is
$$
  \rho = A \epsilon^{1/2}~ \cos[c ~{\rm ln}\epsilon + B]\ \ ,\eqno(2.29)$$
where $c=\sqrt{3+16\psi_0}/2$ with $\psi_0=\psi(\pi/2)$ and
$A,B$ are integration constants.

We numerically found solutions that are consistent with the above asymptotic
limits using a Runge-Kutta method. We integrated equations (2.24) starting
at a point slightly to the right of $-\pi/2$ with the initial condition
that $\rho=+1$  and $\psi=0$. We were then free to vary
the initial derivatives of $\rho(w)$ and $\psi(w)$ giving a two
parameter family of solutions. Typical solutions are shown in
Figs.~3 -- 4.  Note that $\rho$ approaches zero in an oscillatory
manner at the right end point, consistent with the above discussion.

We can take the numerically generated solutions to (2.24) in terms
of $w$, and using (2.21a) and  (2.22a) convert them into space-time
solutions with $r$-$t$ dependence. For very early and
late  times these solutions propagate undistorted keeping their
shape. We may substitute these solutions into the integrand of
(2.18) and obtain an energy
density profile. Fig.~5 shows this profile for a sequence of times.
The energy density forms a localized shell which moves undistorted
in a soliton-like manner at early and late times. This early and
late time behavior can be understood since our solutions depend only
on $z_0$ given by (2.10a). For example, for $t>>1$ and $r$ near
$t$, $z_0$ approaches $t-r$, which shows that our solutions
move undistorted near the speed of light. The behavior of the
energy shell as a function of time is consistent with the
observation of Coleman and Smarr that the radius of gyration,
$\bar r(t)$, for a shell of pure glue must obey $\bar r(t)^2 = t^2 + r_0^2$
where $r_0$ is a constant.$^{16}$

We will see in Section V that only the solutions with $\psi\ne0$ have
interesting
topological properties. However, before studying the topological
structure of these solutions,  we exhibit in the next section
a class of previously discovered solutions whose topological
properties we will also investigate.
\goodbreak
\bigskip
\noindent{\bf III.\quad THE SOLUTIONS OF L\"USCHER AND SCHECHTER}
\medskip

There are other classical solutions in pure $SU(2)$ gauge theory
discovered by L\"uscher$^{10}$ and Schechter.$^{11}$
We will show that these solutions can also be described as
moving spherical
shells of energy. Motivated by the conformal invariance of Euclidean
solutions, L\"uscher and Schechter constructed an $SO(4)$ symmetric
Minkowski solution,
where the $SO(4)$ is a subgroup of the (3+1)-dimensional conformal
group $SO(4,2)$. This $SO(4)$ contains the rotation group, so their solution
can be cast in the spherical ansatz. Converting their solution
to the notation of the previous section we find that $a_\mu$,
$\alpha$ and $\beta$ are expressed in terms of a single function
$q(\tau)$ as
$$\eqalign{
  a_\mu &= -q(\tau)~\partial_\mu w  \ ,\cr
  \alpha &= q(\tau)~ \sin w ~\cos w  \ , \cr
  \beta &=-(1 + q(\tau)\cos^2 w)\ , \cr}\eqno(3.1)$$
where $\mu=t,r$.
Expressing $\rho$ and $\psi$, which now depend on both $\tau$
and $w$,  in terms of (3.1) we find
$$\eqalignno{
  \rho^2(w,\tau) &= 1 + q(q+2)\cos^2 w \ ,&(3.2\hbox{a})\cr
  \psi(w,\tau)   &= {1\over 2} ~ \dot q \cos^2 w \ , &(3.2\hbox{b})}$$
where the dot denotes differentiation with respect to $\tau$.
By substituting (3.1) into (2.10) or equivalently and
more simply substituting (3.2) into (3.23),
it can be seen that $q(\tau)$ obeys:
$$\ddot q + 2q(q+1)(q+2)=0 \ .\eqno(3.3)$$
Note that (3.3) is the equation for an an-harmonic oscillator with the
potential $U={1\over 2} q^2\hbox{(q+2)}^2$. The solutions of (3.3)
may be characterized by the ``energy'' $\varepsilon$ of the
associated mechanical problem,
$$   \varepsilon = {1\over 2} \dot q^2 + U(q)\ . \eqno(3.4)$$
There are two classes of solutions depending on whether $\varepsilon$
is smaller or larger than $1/2$, the barrier height of $U(q)$ at
the unstable point $q=-1$:
$$\eqalign{
  q(\tau) &= -1 \pm (1+\sqrt{2\varepsilon})^{1/2} {\rm dn}\big[
  (1+\sqrt{2\varepsilon})^{1/2}(\tau-\tau_0);k_1\big] \cr
  k^2_1 &=2\sqrt{2 \varepsilon}/(1+\sqrt{2\varepsilon});\qquad \varepsilon
  \le 1/2\ \ ,}\eqno(3.5)$$
and
$$\eqalign{
  q(\tau) &= -1 + (1+\sqrt{2\varepsilon})^{1/2} {\rm cn}\big[
  (8\varepsilon)^{1/4}(\tau-\tau_0);k_2\big] \cr
  k^2_2 &=(1+\sqrt{2\varepsilon})/(2\sqrt{2 \varepsilon});\qquad
  \varepsilon
  > 1/2\ \ ,}\eqno(3.6)$$
where ${\rm dn}$ and ${\rm cn}$ are the Jacobi elliptic functions
and $\tau_0$ is an arbitrary parameter. For $\varepsilon<1/2$, the
``particle'' can be trapped in either well, which corresponds to
the two forms for $q(\tau)$ in (3.5), while there is only one solution
when $\varepsilon>1/2$. The parameter $\tau_0$ corresponds to the
time at which the particle moving in the potential $U(q)$ with
energy $\varepsilon$ is at a turning point.

Note that the solutions (3.5) and (3.6) represent bound motion
in a potential and are therefore periodic in the variable $\tau$.
The period depends upon $\varepsilon$ and in general  is not
$2\pi$ divided by an integer. Therefore $q(\pi)$ does not match
$q(-\pi)$ and, the fields $\rho(w,\tau)$ and $\psi(w,\tau)$ are
generally discontinuous along the line on the hyperboloid $\tau=\pm\pi$.
This line, however, lies outside the physical region $H^{+}$.
Only for special discrete values of $\varepsilon$ for which the period
is $2\pi$ divided by an integer will $\rho$ and $\psi$ match along
the $\tau=\pm\pi$ line and therefore be continuous on the whole
hyperboloid.

As was shown in Ref.~[10], the L\"uscher-Schechter solutions have
finite Minkowski action. The solutions of Section II, however,
have infinite action.

Like the solutions of Section II, the solutions described here also
give spherically symmetric waves of localized energy density.
It can be shown that the energy density is independent of $\tau_0$
and depends only on the parameter $\varepsilon$. Fig.~6
shows some $r$-profiles of the energy density for a sequence
of times. Again in the distant past the solution propagates
undistorted in a soliton-like manner with the energy density
localized in a spherical shell which is collapsing at near the
speed of light. The shell becomes small, distorts, and bounces back
producing an expanding shell. As the shell expands it leaves the
region of space behind it in a pure gauge configuration.
\goodbreak
\bigskip
\noindent{\bf IV.\quad SYMMETRY TRANSFORMATIONS OF THE SOLUTIONS}
\medskip
\nobreak
\def\SOh{SO(2,1)_{\scriptscriptstyle H}}
\def\SOc{SO(2,1)_{\scriptscriptstyle C}}

In this section we show how to construct families of solutions by
acting with symmetry transformations on the solutions we have
described in the previous two sections.
The reader who is primarily interested in the topological properties
of the solutions can skip to the next section.

The fields $\rho$ and $\psi$ which satisfy (2.20) live on the
hyperboloid where each point can be labeled by $z=(z_0,z_1,z_2)$
with $z_0^2 - z_1^2 - z_2^2 = -1$. This applies equally well to the
solutions we found in Section II or the L\"uscher-Schechter solutions
discussed in the previous section. Any solution to (2.20) gives a $\rho(z)$
and a $\psi(z)$ from which we can generate new solutions
$\rho(\Lambda z)$ and  $\psi(\Lambda z)$ where $\Lambda$ is an
element of $SO(2,1)$. Note that this is possible because the transformation
$\Lambda$
maintains the form of the metric $ds^2=d z_0^2 - d z_1^2 - d z_2^2$,
that is the transformation is an isometry. Thus, in general, for each
solution to (2.20) there is actually a three-parameter family of solutions.
We call this $SO(2,1)$ which acts directly on the hyperboloid, $\SOh$.
There is a very closely related $SO(2,1)$ associated with the
conformal group in (3+1) dimensions. To understand this $SO(2,1)$ recall
that pure Yang-Mills theory in (3+1)-dimensions is invariant under the $15$
parameter conformal group $SO(4,2)$. Working in a spherical ansatz, which
entails picking an origin, reduces the symmetry to $SO(3)\times SO(2,1)$.  The
$SO(3)$ is that of spatial rotations whose action on the gauge field $A_\mu$,
given by (2.2), is equivalent to a constant gauge transformation of the form
(2.8).  Since $\rho$ and $\psi$ are gauge invariant functions of $r$ and $t$,
this $SO(3)$ leaves $\rho$ and
$\psi$ unchanged. We call the $SO(2,1)$ subgroup of $SO(4,2)$,
$\SOc$. As we will see $\SOc$ is composed of time translations,
dilatations and special conformal transformations associated with
time translations. We now discuss these two $SO(2,1)$'s and their
relationship.
\goodbreak
\bigskip
\noindent${\bf \SOh}$:
\medskip
\nobreak
The $\SOh$ is the isometry group of the hyperboloid. It takes $\rho(z)$,
$\psi(z)$ into $\rho(\Lambda z)$, $\psi(\Lambda z)$. Note that although  $w$
 and $\tau$  cover the hyperboloid, the
coordinates $r$ and $t$ do not. From (2.21) we see that
$z_0+z_2=1/r$, so  $r$ and $t$ only cover the half of the hyperboloid,
$H^+$, on which $z_0+z_2>0$; see Fig.~1. For a fixed $r$ and
$t$ we determine a
$z \in H^+$ and our new solution at $r$,$t$  is $\rho(\Lambda z)$,
$\psi(\Lambda z)$ even if $\Lambda z$ is not in $H^+$. Thus the
$\SOh$ acts on the whole hyperboloid, even where $r$ and $t$ are not
defined, and in this sense is not a space-time symmetry.

We will decompose $\SOh$ into three one-parameter transformations which
together generate the full $\SOh$. The obvious choice for the three
transformations would be those that leave $z=(1,0,0)$, $z=(0,1,0)$
and $z=(0,0,1)$ invariant. However to see the connection with $SO(2,1)_C$ it
is more convenient to pick a different set.  The first is
\def\LamT{\Lambda_{\scriptscriptstyle T}}
$$  \LamT(c)=
  \left(\matrix{ 1+{1\over2}c^2 & c &{1\over2}c^2\cr\noalign{\vskip 0.2cm}
  c & 1 & c \cr\noalign{\vskip 0.2cm}
  -{1\over2}c^2 & -c &1-{1\over2}c^2 \cr\noalign{\vskip 0.2cm}}\right)
  \ \ ,\eqno(4.1)$$
which depends on $-\infty<c<\infty$. Note that $z_0+z_2$ is invariant
under this transformation so $H^+\to H^+$. If we use (2.21) to
discover the action of the transformation on $r$ and $t$ we find
$$  r \to r \ ; \qquad t \to t+c \ \ ,
\eqno(4.2)$$
so clearly $\LamT$ generates time translations. The second transformation
is
\def\LamD{\Lambda_{\scriptscriptstyle D}}
$$
  \LamD(\alpha)=
  \left( \matrix{ \cosh\alpha & 0 & -\sinh\alpha\cr\noalign{\vskip 0.2cm}
  0 & 1 & 0 \cr\noalign{\vskip 0.2cm}
  -\sinh\alpha & 0 &\cosh\alpha \cr\noalign{\vskip 0.2cm}}\right) \ \ ,
\eqno(4.3)$$
with $-\infty<\alpha<\infty$. Here $z_0+z_2 \to e^{-\alpha}(z_0+z_2)$
so the sign of $z_0+z_2$ is invariant and $H^+ \to H^+$. Acting on
$r$ and $t$ this gives
$$
  r \to e^{-\alpha}~r \ ; \qquad t \to e^{-\alpha}~t \ ,
\eqno(4.4)$$
which is the dilatation transformation. The two transformations
(4.1) and (4.3) together form a two-parameter non-semisimple
subgroup of $\SOh$ which maps $H^+$ into $H^+$.

The third transformation is
\def\LamS{\Lambda_{\scriptscriptstyle S}}
$$
     \LamS(d)=
     \left( \matrix{
     1+{1\over2}d^2 & -d & -{1\over2}d^2\cr\noalign{\vskip 0.2cm}
     -d & 1 & d \cr\noalign{\vskip 0.2cm}
     {1\over2}d^2 & -d &1-{1\over2}d^2 \cr\noalign{\vskip 0.2cm}}\right)
     \ \ ,\eqno(4.5)$$
with $-\infty<d<\infty$. For a fixed $d$, if $(1-dt)^2-d^2r^2>0$
then the $z$ associated with $r,t$ is mapped under (4.5) into
$H^+$ and we have
$$\eqalign{
  r & \to {r\over (1-dt)^2-d^2r^2} \ , \cr\noalign{\vskip 0.2cm}
  t & \to {t+d(r^2-t^2) \over (1-dt)^2-d^2r^2} \ \ .\cr}\eqno(4.6)$$
However if $(1-dt)^2-d^2r^2<0$, then the $z$ associated with
$r,t$ is mapped into $H^-$, the half of the hyperboloid on which
$z_0+z_2<0$. Note that for any $d\ne 0$, there are always points
$z\in H^+$ which are mapped into $H^-$. We will have more to say
about the transformation (4.6) when we discuss $\SOc$. Also note
that if we define the discrete transformation $I_0(z_0,z_1,z_2)
\equiv (-z_0,z_1,z_2)$ then we see that
$$
  \LamS(d) = I_0~\LamT(d)~I_0 \ \ .\eqno(4.7)$$

In principle any solution of (2.20) which lies on the whole
hyperboloid corresponds to a three parameter family of solutions
generated by $\SOh$. The solutions introduced in Section II
depend only on $z_0$ and not on the ratio $z_1/z_2$. Thus the
transformations which leave $z_0$ invariant do not, in this
case, generate new solutions and there is just a two parameter
family of solutions. To see this in terms of the transformations
just introduced, note that $\LamS(d)$ given by (4.5) can be
written as the product of a rotation about $(1,0,0)$ times a
dilatation (4.3) times a time-translation (4.1),
$$
  \LamS(d) = R_0(\beta)~\LamD(\ln(1+d^2))~\LamT(-{d\over 1+d^2}) \ \
.\eqno(4.8)$$
Here $R_0(\beta)$ is a rotation about $z=(1,0,0)$ by the angle
$\beta$ where $\cos\beta=(1-d^2)/(1+d^2)$ and $\sin\beta =
2d/(1+d^2)$. Since $R_0(\beta)$ leaves the solutions of Section II
invariant we see that $\LamS(d)$ acting on the solutions can be
expressed in terms of the transformations $\LamD$ and $\LamT$.

Except at special values of $\varepsilon$, the L\"uscher-Schechter
solutions are discontinuous on the hyperboloid along $\tau=\pm\pi$.
Recall that $\tau=\pm \pi$ is in $H^-$. The transformations $\LamT(c)$
and $\LamD(\alpha)$ take $H^+\to H^+$ so these transformations do
not bring the cut into $H^+$ and therefore new solutions, generated
by $\LamT(c)$ and $\LamD(\alpha)$, are continuous. The same can not be
said of $\LamS(d)$ since by (4.8) we see that the $\tau=\pm\pi$
line is rotated into $H^+$ by $R_0$ for any $\beta\ne0$. Thus the
transformation $\LamS(d)$ acting on the solutions of Section III
does not generate new solutions. (We do not call a function a solution
to a differential equation if it solves the equation everywhere except
along a line where the function is discontinuous.) The L\"uscher-Schechter
solutions can be parametrized by $\varepsilon$ and $\tau_0$ together with
the parameters $\alpha$ and $d$ associated with dilatations
and time-translations. For the values of $\varepsilon$
for which $\rho$ and $\psi$ match at $\tau=\pm \pi$ and there is no cut,
one may think that $\LamS(d)$ generates new solutions. However,
these solutions can be obtained from the original solutions
by a shift of $\tau_0$.

Because the L\"uscher-Schechter solutions are discontinuous along
$\tau=\pm\pi$ on the hyperboloid, the transformation $R_0(\beta)$ under
which $\tau\to\tau+\beta$ can not be used to generate new solutions
since the transformation brings the cut into $H^+$. However, the cut exists
only because we insist on identifying $\tau=\pi$ with
$\tau=-\pi$. If we look at the differential equations (2.23) we
can imagine producing a solution for $-\infty<\tau<\infty$. In
this sense we would be solving on a multiply covered hyperboloid and there
would be no cuts. The transformation $R_0(\beta)$ would
take $\rho(\tau,w)$, $\psi(\tau,w)$ to $\rho(\tau+\beta,w)$,
$\psi(\tau+\beta,w)$.  However, shifting $\tau$ is equivalent to
shifting $\tau_0$ (see (3.5) and (3.6)), so actually no new solutions
are introduced by adopting this point of view.
\goodbreak
\bigskip
\noindent${\bf \SOc}$
\medskip
As we discussed in the beginning of this section, the full
$SO(4,2)$ conformal group in (3+1) dimensions has an $\SOc$
subgroup which commutes with the $SO(3)$ of rotations about
${\bf x}={\bf 0}$. This group can be decomposed into three
continuous transformations which generate the full group:
\medskip
\item{(i)}
time translations: $t\to t+c$, $x^i\to x^i$.
Acting on functions that depend only on $r$ and $t$ this is
equivalent to (4.2).
\medskip
\item{(ii)}
dilatations: $t\to e^{-\alpha} t$, $x^i\to
e^{-\alpha} x^i$. Again acting on functions that depend only on
$r$ and $t$ on this is equivalent to (4.4).

\noindent
Thirdly, we have the special conformal transformation associated
with the time translation. This is generated by an inversion,
$x^\mu \to x^\mu/x^2$, followed by a time translation, followed
by another inversion.  The net transformation is
\vfil\eject
\item{(iii)}
special conformal:

$$\eqalign{
  x^i &\to {x^i\over (1-dt)^2-d^2r^2} \ ,\cr\noalign{\vskip 0.2cm}
  t &\to {t+d(r^2-t^2)\over (1-dt)^2-d^2r^2} \ .
}\eqno(4.9)$$
\smallskip\noindent
Note that along the light cone, $(1-dt)^2-d^2r^2=0$, the denominators
in (4.9) vanish and the transformed coordinates blow up. As you
cross this light cone and $(1-dt)^2-d^2r^2$ changes sign, the
transformed coordinates change discontinuously from say $-\infty$
to $+\infty$. This can induce discontinuities in the transformed
solution unless the original solution is particularly well behaved
as $x^i\to\pm\infty$ and $t\to\pm\infty$.

Acting on functions which depend only on $r$ and $t$, the special
conformal transformation given above is
$$\eqalign{
  r & \to {r\over |(1-dt)^2-d^2r^2|}\ , \cr\noalign{\vskip 0.2cm}
  t & \to {t+d(r^2-t^2) \over (1-dt)^2-d^2r^2} \ ,
}\eqno(4.10)$$
which is the same as (4.6) only when
                                    $(1-dt)^2-d^2r^2>0$. The
$\SOc$ is a space-time symmetry. It takes fields defined on
${\bf x},t$ and generates new fields on ${\bf x},t$ using the
old values of the fields. This is distinct from $\SOh$ which
can take values of the fields on $H^-$ and bring them into
$H^+$. For space-time functions $\rho(r,t)$, $\psi(r,t)$ we
can still understand how $\SOc$ acts by looking at the hyperboloid.
For each $r,t$ there is a $z \in H^+$, so given $\rho(r,t)$ and $\psi(r,t)$
we know $\rho(z)$ and $\psi(z)$ on $H^+$. Now define
new fields $\tilde\rho(z)$ and $\tilde\psi(z)$ by
$$
  \tilde\rho(z)=\cases{\rho(z), & if $z\in H^+$\cr
                       \rho(-z), & if $z\in H^-$\cr}\eqno(4.11)$$
and similarly
$$  \tilde\psi(z)=\cases{\psi(z), & if $z\in H^+$\cr
                       \psi(-z), & if $z\in H^-$\ \ .\cr}\eqno(4.12)$$
Thus $\tilde\rho$ and $\tilde\psi$ are defined everywhere on the hyperboloid.
We now define $\rho'(z)=\tilde\rho(\Lambda z)$ and
$\psi'(z)=\tilde\psi(\Lambda z)$ with $\Lambda\in\SOh$ from
which we can infer $\rho'(r,t)$ and $\psi'(r,t)$. For
$\Lambda=\LamT(c)$ of (4.1) it is immediate that $\rho'(r,t)=
\rho(r,t+c)$ and $\psi'(r,t)=\psi(r,t+c)$; for $\Lambda=
\LamD(\alpha)$ of (4.3) it is immediate that $\rho'(r,t)=
\rho(e^{-\alpha}r,e^{-\alpha}t)$ and $\psi'(r,t)=
\psi(e^{-\alpha}r,e^{-\alpha}t)$. For $\Lambda=\LamS(d)$
of (4.5), for which $\Lambda z$ can be in $H^-$, we use
the fact that $z\to-z$ is equivalent to $r\to-r$ in (2.21)
to show that $\rho'(r,t)=\rho(r',t')$ and $\psi'(r,t)=\psi(r',t')$
with $r'$ and $t'$ given by (4.10).

In short the action of $\SOh$ and $\SOc$ can be summarized as follows.
For a solution which is defined on the whole hyperboloid, $\SOh$
defines a new solution $\rho(\Lambda z)$, $\psi(\Lambda z)$ with
$\Lambda \in \SOh$. The $\SOc$ is a space-time symmetry and only
refers to $\rho(z)$ and $\psi(z)$ defined on $z\in H^+$. To find
the action of $\SOc$ you discard $\rho(z)$ and $\psi(z)$ for
$z\in H^-$ and replace them with $\rho(-z)$ and $\psi(-z)$ giving
$\tilde\rho$ and $\tilde\psi$ defined on the whole hyperboloid
(see (4.11) and (4.12)). The $\SOc$ gives new solutions
$\rho'(z)=\tilde\rho(\Lambda z)$, $\psi'(z)=\tilde\psi(\Lambda z)$
with $\Lambda\in\SOh$. For time translations and dilatations the
two $SO(2,1)$'s act in precisely the same manner. However, for $\Lambda$'s
given by (4.5) they are different transformations. For a fixed $d$
and $r$,$t$ such that $(1-dt)^2-d^2r^2>0$, the two transformations
agree. However for $(1-dt)^2-d^2r^2<0$ they disagree.

In using (4.11) and (4.12) we should note that the boundary
of $H^+$ and $H^-$ is the intersection of the plane $z_0+z_2=0$
with the hyperboloid. For $\tilde\rho(z)$ and $\tilde\psi(z)$ to
be continuous on the hyperboloid we require $\rho(z_0,z_1,-z_0)=
\rho(-z_0,-z_1,z_0)$ and $\psi(z_0,z_1,-z_0)=\psi(-z_0,-z_1,z_0)$.
If these conditions are not met, then $\tilde\rho$ and $\tilde\psi$
are discontinuous along $z_0+z_2=0$. In this case $\tilde\rho(\LamS z)$
and $\tilde\psi(\LamS z)$
                         viewed as functions of $r$,$t$ are
discontinuous and the special conformal transformation does not produce
new solutions. The solutions of Section II are functions of $z_0$ only,
but these functions are not even, so the special conformal
transformation acting on these solutions produces ripped functions.
Similarly no new solutions are produced by acting with special
conformal transformations on the L\"uscher-Schechter solutions.
\goodbreak
\bigskip
\noindent{\bf V.\quad TOPOLOGICAL PROPERTIES OF CLASSICAL SOLUTIONS}
\medskip
\noindent
In Section II we found a new class of Minkowski space solutions
to the Yang-Mills equations and in Section III we described another
class of solutions discovered by L\"uscher and Schechter.
Both of these classes have the following general properties. At a large
negative time, $-T_0$, the energy density of the solution is localized
in a thin spherical shell with radius of order $T_0$, and the shell
is collapsing. Imagine a 2-sphere of radius $R$, which we call $S_R$,
which is concentric with the energy shell. For $T_0>>R$ the energy density
of the classical solution is initially
localized far outside of $S_R$ so the region inside this
sphere is very close to pure gauge. At a time $t\sim -R$ the energy front
reaches $S_R$. Energy flows in,
bounces back at time $t \sim 0$, and then leaves the region at time
$t\sim R$.
At large positive time, $T_0$, the fields inside $S_R$
are once again pure gauge. But this final vacuum configuration need not
coincide with the initial one.
\goodbreak
\bigskip
\noindent {\bf The Local Winding Number Change}
\medskip
\medskip
In the $A_0=0$ gauge, the initial and final vacuum configurations
inside $S_R$ are given by pure gauges of the form (2.7).
Such configurations inside the sphere can be characterized by
the quantity
$$
  \nu(R) = {1\over 24\pi^2}\int_R d^3{\bf x} ~\epsilon^{ijk}
  {\rm tr}\big[ \big(U^\dagger \partial_i U\big)
  \big(U^\dagger \partial_j U\big)\big(U^\dagger \partial_k U\big)
  \big] \ \ ,\eqno(5.1)$$
where the integration is over  the interior of $S_R$.
We will refer to this expression as the local winding
number. A pure gauge configuration in the interior of
$S_R$ with $U=1$ (or any constant element of $SU(2)$)
on the boundary defines a map from compactified three
space -- the three sphere -- into  the gauge group. In this case
$\nu(R)$ is an integer, the winding number, which characterizes the homotopy
class of the map. If $U$ is not required to equal unity
(or a constant element of $SU(2)$) on $S_R$, $\nu(R)$ will not in
general be an integer. Furthermore if we do not restrict $U$ on
$S_R$, then $\nu(R)$ will not be additive on the products of
two $U$'s.

The solutions which we consider excite the interior of $S_R$ only
for $-R~ \lapp ~t~ \lapp ~R$. Before and after this period the field
inside the sphere
is pure gauge. We are interested in calculating the winding number
change of these
                                       pure gauge configurations.
Working in the $A_0=0$ gauge we are free to make time-independent
gauge transformations. We use this freedom to set $U=1$ at points
inside of $S_R$ at very early times. Therefore $\nu(R)=0$ at these
early times. Inside of $S_R$ there is no gauge freedom left and the
form of $U$
inside of $S_R$ at late times is now determined by the classical
solutions. The quantity we are after is (5.1)  evaluated with
$U$ determined by the late time behavior of the solutions.

\def\UE{U_{\scriptscriptstyle E}}
\def\UL{U_{\scriptscriptstyle L}}
The prescription we have just given for determining the local
winding number change associated with a solution is in fact gauge
invariant. Consider a solution $A_\mu({\bf x},t)$ which has the
property that at very early and very late times, $A_\mu$ is pure
gauge for $|{\bf x}| < R$. Go into the $A_0=0$ gauge. At very
early times $A_j={i\over g}\UE \partial_j \UE^\dagger$ and at very
late times $A_j={i\over g}\UL \partial_j \UL^\dagger$ for
$|{\bf x}|<R$. Inside of $S_R$ make a gauge transformation by
$\UE^\dagger$ so that in the far past $A_j=0$. Now at late times
$A_j={i\over g}U \partial_j U^\dagger$ with $U=\UE^\dagger \UL$.
We then evaluate $\nu(R)$ given by (5.1) for this $U$. If we had
started with a gauge transform of $A_\mu$, say $A'_\mu$, and then
passed to the $A'_0=0$ gauge we would have discovered that $\UE$ goes
into  $V \UE$ and $\UL$ goes into $V \UL$ where $V$ is time independent.
Thus $U=\UE^\dagger \UL$ is gauge invariant. We are evaluating the
local winding number of the difference between $\UE$ and $\UL$
which is $\UE^\dagger \UL$. Note that the local winding number
of $\UE^\dagger \UL$ is {\it not} the difference
between the local winding numbers of $\UL$ and $\UE$ since
$\nu(R)$ is not in general additive.

We now return to the spherical ansatz where $U$ has the form (2.8).
Substituting this form into (5.1)  gives the following expression
for the local winding number inside of $S_R$ in terms of the gauge
function $\Omega(r)$ which characterizes the late time pure gauge
inside of $S_R$:
$$
  \nu(R) = {1\over 2\pi}\int_0^R dr~ {d\Omega(r) \over dr}~\left[
  1 - \cos\Omega(r) \right] \ \ .\eqno(5.2)$$

Note that the expression for the local winding number (5.2) which
arises from a (3+1)-dimensional discussion is different from the expression
for the local winding number, $\omega$, which one might
naively infer from the (1+1)-dimensional $U(1)$ gauge theory vacuum
given by (2.9),
 $$
  \omega(R) = {1\over 2\pi} \int_0^R dr~{d\Omega(r) \over dr} \ \ .
\eqno(5.3)$$
The right hand sides of (5.2) and (5.3) coincide, however,
if $\Omega(r)$ is an integer multiple of $2\pi$ at $r=0$ and $r=R$.
This illustrates that the definition (5.1) of the local winding
number is somewhat arbitrary.
Indeed, you might pick $\nu(R)$ to be any function of $U$
which coincides with (5.1) when $U$ is restricted to a constant
on $S_R$.

We now express the local winding number (5.2) of the final vacuum
inside of $S_R$ as $t \to \infty$ in terms of  finite-energy classical
solutions. Using
(2.9) and (2.11) we see that $\varphi(r,t)\to\Omega(r)$ as
$t\to +\infty$ for any fixed $r$. Thus, (5.2) gives
$$
  \nu(R) = {1\over 2\pi}\bigl[ ~\varphi(r,\infty) -
  \sin\varphi(r,\infty)~\bigr]\bigg\vert_{r=0}^{r=R} \ \ \eqno(5.4)$$
Now the finite energy condition (2.19a) gives $\rho(0,t)=1$
for all $t$. In fact, it can be shown that all finite energy
solutions obey $\partial_r \psi(r,t) \big\vert_{r=0} = 0$,
from which (2.15) implies $\partial_t
\varphi(0,t)=0$. Thus we have that at $r=0$, $\varphi$ stays locked down
at its initial value, which we have taken to be zero, so $\varphi(0,t)=0$.
Recall from our discussion following (2.9) that
we must require $\Omega(r=0)$ to be equal to zero for pure gauges in
(3+1)-dimensions.
We now see that this condition is dynamically imposed on all finite
energy solutions. Therefore equation (5.4) can now be written as
$$
  \nu(R) = {1\over 2\pi}\bigl[~ \varphi(R,\infty) -
  \sin\varphi(R,\infty)~\bigr] \ \ .\eqno(5.5)$$
Note that $\nu(R)$ is intimately related to the change in the phase of
the field $\chi$ at $R$ produced by the passing energy front.

We are interested in evaluating $\varphi(R,\infty)$ in the $a_0=0$
gauge. In the late-time limit, $t \to \infty$, the fields are pure
gauge for any finite $r$, so $\partial_r\varphi(r,\infty) = a_1(r,\infty)$.
Thus
$$\eqalign{
  \varphi(R,\infty) &= \int_0^R dr~ a_1(r,\infty) \cr
  &=\int_{-\infty}^\infty dt \int_0^R dr~ \partial_0 a_1(r,t) \cr
  &=-{1\over 2} \int_{-\infty}^\infty dt \int_0^R dr~
  \epsilon^{\mu\nu}f_{\mu\nu} \ \ ,\cr}\eqno(5.6)$$
where the last expression is manifestly gauge invariant. Using
(2.13) we have
$$
  \varphi(R,\infty) = -2 \int_{-\infty}^\infty dt \int_0^R dr~
  {\psi(r,t) \over r^2} \ \ ,\eqno(5.7)$$
which can be expressed in terms of the hyperboloid variables $w$
and $\tau$ as
$$
  \varphi(R,\infty) = -2 \int_{H_R}dw d\tau~
  {\psi(w,\tau) \over \cos^2 w} \ \ ,\eqno(5.8)$$
where $H_R$ is the part of $H^+$ for which $r \le R$. From (2.21)
we see that the fixed $R$ contour is given by the intersection of
the plane $z_0+z_2 = 1/R$ with the hyperboloid.  In terms of $w$ and
$\tau$, from (2.22), we have that at fixed $R$, $w$ and $\tau$
obey
$$
  \cos\tau = {1\over R}\cos w + \sin w \ \ .\eqno(5.9)$$
The region $H_R$ is shown in Fig.~7.  We can now write (5.8) as
$$
  \varphi(R,\infty) = -2 \int_{-\pi/2}^{w_{\rm max}}dw~
  \int_{-\tau_+(w,R)}^{\tau_+(w,R)} d\tau~~
  {\psi(w,\tau) \over \cos^2 w} \ \ ,\eqno(5.10)$$
where $\tau_+(w,R)$ is the positive value of $\tau$ which solves
(5.9) for fixed $R$ and $w$, and $w_{\rm max}$ obeys
$$
  \sin w_{\rm max} = {R^2-1 \over R^2+1} \ \ .\eqno(5.11)$$
It is now straightforward to evaluate (5.10) for any solution
in the spherical ansatz.

\noindent
{\it The Local Winding Number Change Produced by the
Solutions of Section II}
\hfill\vskip0.1cm

Here we discuss $\nu(R)$ given by (5.5) for the solutions of
Section II. There is no reason to expect $\nu(R)$ to have an
integer value at any $R$ and in fact $\nu(R)$ diverges     for
the solutions of Section II as $R\to\infty$. In evaluating (5.5)
we use (5.10) and note that in this case $\psi$ is independent
of $\tau$, so
$$
  \varphi(R,\infty) = -4 \int_{-\pi/2}^{w_{\rm max}} dw~~ \tau_+(w,R)~
  {\psi(w) \over \cos^2 w} \ \ .
\eqno(5.12)$$
For $w$ near $-\pi/2$, the integrand is integrable because
$\psi(w)$ goes to zero fast enough; see (2.28b).
For finite $R$, $w_{\rm max}$ is less than $\pi/2$ and accordingly
$\psi(R,\infty)$ and $\nu(R)$ are finite. In Fig.~8 we show
$\nu(R)$ versus $R$
for a typical solution.  Note that $\nu(R)$ does not
appear to
have a limit as $R$ goes to infinity. This is indeed the case which
can be seen from (5.12). As $R\to\infty$, $w_{\rm max} \to \pi/2$ and
$\tau_+(w,R) \to \pi/2 - w$. Since $\psi(\pi/2)\ne 0$ the integral
diverges at its upper limit.

\noindent
{\it The Local Winding Number Change Produced by the
L\"uscher-Schechter Solutions}
\hfill\vskip0.1cm

The local winding number $\nu(R)$ of a L\"uscher-Schechter solution
approaches a fixed value as $R\to\infty$. To see this we again
evaluate (5.5) using (5.10) for fixed $R$. However in this case
from (3.2b),
$\psi(w,\tau) = {1\over 2} \dot q(\tau)\cos^2 w$ so
$$
  \varphi(R,\infty) = \int_{-\pi/2}^{w_{\rm max}} dw~\big\{
  q[-\tau_+(w,R)] - q[\tau_+(w,R)] ~\big\} \ \ .\eqno(5.13)$$
Recall from the discussion in Section III that $q(\tau)$ can be
viewed as the coordinate of a particle moving in a potential
$U(q)$ with ``energy'' $\varepsilon$ as expressed by (3.4).
Besides $\varepsilon$, the solutions are characterized by
$\tau_0$, the ``time'' when the particle is at a turning point
of $U(q)$. If $\tau_0=0$, then $q(\tau)$ is an even function
and (5.13) vanishes for all $R$. To have non-trivial topological
properties we must look at $\tau_0 \ne 0$. In Fig.~9 we give
examples of $\nu(R)$ versus $R$ for $\tau_0=1$ and
$\varepsilon=0.7$ and $2.0$. Note that as $R\to\infty$, $\nu(R)$ appears
to approach a limit. To obtain a simple expression for this
limiting value, note as we mentioned just before, that  as
$R\to \infty$, $w_{\rm max}\to \pi/2$ and $\tau_+(w,R)\to
\pi/2 -w$. Thus as $R$ goes to infinity, (5.13) gives
$$
  \lim_{R\to\infty} \varphi(R,\infty) = \int_0^\pi d\tau~
  \big[ q(-\tau) - q(\tau) \big] \ \ .\eqno(5.14)$$
In Figs.~10 and 11 we show the asymptotic
winding number $\nu \equiv \lim_{R\to\infty} \nu(R)$ which is obtained
from (5.14). We plot both $\nu$ versus $\varepsilon$ for fixed $\tau_0$
and $\nu$ versus $\tau_0$ for fixed $\varepsilon$.  In the latter case the
asymptotic winding number is periodic in $\tau_0$.  This period is exactly the
period of a particle moving in the potential $U(q)$ with energy
$\varepsilon$.
\goodbreak
\bigskip
\noindent{\bf The Local Topological Charge}
\medskip
\nobreak
We first discuss the local topological charge of any solution
in the spherical ansatz. We consider the local topological
charge defined as
$$
  Q(R) = {g^2 \over 16 \pi^2}
  \ \int_{-\infty}^{\infty}dt~ \int_R d^3x~ {1\over
   2}\epsilon^{\mu\nu\alpha\beta}{\rm tr} (F_{\mu \nu} F_{\alpha \beta})
   \ \ ,
\eqno(5.15)$$
where the spatial integration is over the interior of the sphere
of radius $R$. This quantity is manifestly gauge invariant. The topological
charge, $Q$, given in (1.1) is the
limit of (5.15) as $R\to\infty$. Since the topological charge density
is the divergence of a current, the topological charge can be expressed
as a surface integral. In the spherical ansatz we can write the following
expression$^{13,14}$ for $Q(R)$,
$$
  Q(R) = \int_{-\infty}^{\infty}dt~ \int_0^R dr~
  \partial_\mu j^\mu \ \ ,\eqno(5.16)$$
where the two dimensional current $j^\mu$ is
$$
  j^\mu = -{\epsilon^{\mu\nu}\over 2\pi}\left[ a_\nu -
  {\rm Re}~\partial_\nu\chi +
  {1\over 2i}\bigg(\chi^*D_\nu\chi - \chi(D_\nu\chi)^*\bigg)\right] \ \ .
\eqno(5.17)$$
Integrating (5.16) gives
$$
  Q(R) =              \int_0^R dr~j^t\bigg\vert_{t=-\infty}^{t=\infty} +
              \int_{-\infty}^\infty dt~j^r\bigg
  \vert_{r=0}^{r=R} \ \ .\eqno(5.18)$$
The first term we recognize as the change in winding number $\nu(R)$
given by (5.5) and the second term is the net flux through $R$.

For solutions to the equations of motion $\partial_\mu j^\mu$
takes a simple form. In terms of the gauge invariant variables
$\rho$ and $\psi$ we have
$$
  Q(R) = {1\over 2\pi}
  \int_{-\infty}^\infty dt \int_0^R dr~ \left[
  -{2\psi \over r^2} + \partial^\mu\partial_\mu \psi \right] \ \ ,\eqno(5.19)$$
which upon using (5.7) gives
$$
  Q(R) = {1\over 2\pi}\varphi(R,\infty) +
  {\cal F}_R              \ \ ,
\eqno(5.20)$$
where      the first term in this decomposition of $Q(R)$ differs
from $\nu(R)$ by ${1\over 2\pi}\sin[\varphi(R,\infty)]$ and the second
term,
                          ${\cal F}_R$, is defined as follows:
$$
  {\cal F}_R = {1\over 2\pi}\int_{-\infty}^\infty dt \int_0^R dr~
  \partial_\mu(\tilde \eta^{\mu\nu} \partial_\nu \psi) \ \ .
\eqno(5.21)$$
Here  again $\tilde\eta^{\mu\nu}={\rm diag}(-1,+1)$. Since
$ds^2 = (dr^2-dt^2)/r^2 =(dw^2-d\tau^2) /\cos^2 w $
it follows that
$$
  {\cal F}_R = {1\over 2\pi} \int_{H_R} dw d\tau~
  \partial_\alpha(\tilde \eta^{\alpha\beta} \partial_\beta \psi) \ \ ,
\eqno(5.22)$$
where $\alpha,\beta=\tau,w$. We can express
${\cal F}_R$ as an integral along the boundary of $H_R$:
$$
  {\cal F}_R = {1\over 2\pi}\int_{-\pi}^{\pi} d\tau~
  \left[{\partial \psi \over \partial w} +
  {d w_b \over d \tau}~{\partial \psi \over \partial \tau}
   \right]\bigg\vert_{w=w_b(\tau,R)} \ ,
\eqno(5.23)$$
where $w_b(\tau,R)$ is the value of $w$ determined by (5.9)
for fixed $R$ and $\tau$. In deriving (5.23) we have used the
fact that for finite energy solutions ${\partial \psi \over
\partial w}\big\vert_{w=-\pi/2}=0$, so there is no contribution
along the lower $w=-\pi/2$ boundary of $H_R$ as seen in Fig. 7.
                                     We will shortly evaluate
${\cal F}_R$ for the two classes of solutions we have discussed
and obtain $Q(R)$ from (5.20).

We can also obtain a simple expression for $\lim_{R\to\infty}
{\cal F}_R$. Note that as $R\to\infty$,
$w_b \to \pi/2 + \tau$ for $\tau<0$ and $w_b \to \pi/2 - \tau$ for $\tau>0$.
In terms of the variables $w_{\scriptscriptstyle \pm}
\equiv w\pm\tau$, (5.23) implies
$$\eqalign{
  \lim_{R\to\infty}{\cal F}_R &= {1\over 2\pi}
  \int_{-3\pi/2}^{ \pi/2} dw_{\scriptscriptstyle +}~
  {\partial \psi\over \partial w_{\scriptscriptstyle +}}
  \bigg\vert_{w_-=\pi/2} ~-~
  {1\over 2\pi}
  \int_{\pi/2}^{-3\pi/2} dw_{\scriptscriptstyle -}~
  {\partial \psi\over \partial w_{\scriptscriptstyle -}}
  \bigg\vert_{w_+=\pi/2}
  \cr\noalign{\vskip 0.3cm}
  &= {1\over \pi}\psi(w=\pi/2,\tau=0) \ \ ,}\eqno(5.24)$$
where we used the fact that for all finite energy
solutions $\psi(w=-\pi/2, \tau) =0$.
For solutions with the property that
                       $\varphi(R,\infty)$ has a
{\it finite} limit as
$R\to\infty$, (5.10) implies that \break $\psi(w=\pi/2,\tau=0)=0$. In  this
situation we obtain
$$
  Q = {1\over 2\pi}\lim_{R\to\infty} \varphi(R,\infty) \ \ .\eqno(5.25)$$

\vfil\eject
\noindent
{\it The Local Topological Charge of the Solutions of Section II}
\hfill\vskip0.1cm

For the solutions of Section II where $\psi$ depends only
on $w$, we have from (5.23) that
$$
  {\cal F}_R = {1\over 2\pi}\int_{-\pi}^\pi d\tau~
  {\partial \psi \over \partial w}\bigg\vert_{w=w_b(\tau,R)} \ \ .
\eqno(5.26)$$
For a fixed value of $R$ we can evaluate this expression given a
numerically generated solution $\psi(w)$.
To get $Q(R)$ we use (5.20) with
${\cal F}_R$ given by (5.26) and $\varphi(R,\infty)$ given by
(5.12). An example of $Q(R)$ as a function of $R$ is shown in
Fig.~8.  Again $Q(R)$ has no limit as $R\to\infty$ since
$\varphi(R,\infty)$ does not approach a limit.

\noindent
{\it The Local Topological Charge of the L\"uscher-Schechter
Solutions}
\hfill\vskip0.1cm

Here again $\psi(w,\tau) = {1\over 2}\dot q(\tau) \cos^2 w$. Thus
from (5.23)
$$
  {\cal F}_R = -{1\over 4\pi}\int_{-\pi}^\pi d\tau~
  \left[\dot q \sin 2w -  \ddot q \cos^2 w ~{d w_b\over d\tau}\right]
  \bigg\vert_{w=w_b(\tau,R)} \ .
\eqno(5.27)$$
Now if $\tau_0=0$, $q(\tau)$ is an even function and ${\cal F}_R=0$.
Thus for $\tau_0=0$, $Q(R)=0$. For $\tau_0 \ne 0$, in general
${\cal F}_R$ will not vanish. We can evaluate ${\cal F}_R$ for a
given solution and then with the use of (5.20) we can determine
$Q(R)$. Examples are given in Fig.~9.

For the L\"uscher-Schechter solutions, $\varphi(R,\infty)$
has a limit as $R \to \infty$ as can be seen from (5.14).
Thus (5.25) applies and we can obtain $Q$ which we see is not
an integer. Figures 10 and 11 show the  asymptotic
topological charge for typical L\"uscher-Schechter
solutions.

\noindent
{\it Integer Topological Charge?}
\hfill\vskip0.1cm

You may think that $Q=~\lim_{R\to\infty}Q(R)$ should be an integer
because of general topological arguments. However, this is not
so, essentially because finite energy solutions are to
be found at arbitrarily large radii for arbitrarily large times.
The argument which leads to integer values of $Q$ is as follows.
Consider a region of space-time which contains non-zero energy
and imagine that outside this region the energy density is zero.
You can surround the region of space-time by a three dimensional
surface which is topologically $S^3$. On this surface the gauge
field is pure gauge, {\it i.e.} $A_\mu =i/g \ U \partial_\mu U^\dagger$.
Thus we have a map from $S^3$ into the gauge group
which is characterized by an integer. The topological charge,
integrated over the space-time region in question, gives this integer.
For non-zero energy solutions to the equations of motion, it
is not possible to surround the energy density by a three-sphere
in space-time because the energy density is moving out to
infinity at early and late times. We do not expect and do not find
integer values of $Q$.
\goodbreak
\bigskip
\noindent{\bf VI.\quad DISCUSSION}
\medskip
\nobreak
Our ultimate aim is to relate the classical solutions
discussed in this paper to physical processes. In a companion
paper $^9$ we analyze fermion production in the background
of a field which locally changes winding number. We work with a
(1+1)-dimensional analogue. We show that non-integer change in winding
number is associated with quantum mechanical fermion number production
where the change in winding is the expectation of the number of fermions
produced. Therefore we believe that if we could produce a field
configuration like one discussed in this paper, it would be associated with
fermion number violation.
Here we are ignoring the back-reaction that the fermions have on
gauge fields.

We are then led to seek the relationship between the classical solutions
and the gauge boson quanta of the real world. If the classical solution
represents a coherent quantum state, then we need to understand the overlap
of this state with the few particle quantum state of an accelerator beam
if we are going to use our methods to estimate rates for fermion violation
in actual experiments. We plan to pursue this as well as the question of
the relative weight these coherent states have in a high temperature
density matrix which should tell us the relevance of these solutions
to high temperature processes.

We are intrigued by the fact that these solutions change the winding
number of local regions of space by non-integer amounts and create
non-integer topological charge. Using a solution
to the classical field equations you can excite an arbitrary large
region of space and discover that after the energy has dissipated from
the region, the winding number of the region will have changed
by a non-integer amount. The implications of this for the full quantum
theory, we have yet to discover.
\goodbreak
\bigskip
\centerline{\bf ACKNOWLEDGEMENTS}
\medskip
\nobreak
We are grateful to Orlando Alvarez, Michael Crescimanno,
Jeffrey Goldstone, Alan Guth, Roman Jackiw, Ken Johnson
and Valery Rubakov for very useful discussions at different
stages of this work.
\vfill
\eject
\centerline{\bf APPENDIX}
\medskip
\def\winit{w_{\rm init}}
\def\wfin{w_{\rm fin}}

In this appendix we investigate the analytic structure of solutions
(2.24) near the end points $\winit=-\pi/2$ and $\wfin=+\pi/2$.
{}From (2.21) and (2.22) we can see that $\winit$ corresponds
to $r=0$ for fixed time $t$, $\wfin$ corresponds to $r=\infty$
for fixed time $t$, and that $t=\pm\infty$ for fixed $r$ also
corresponds to $\winit$.
Recall from (2.19) that for finite energy solutions  $\rho$ is near
one and $\psi$ is near zero for $w$ near $\winit$.  Writing
$w=\winit
+ \epsilon$, from the differential equations (2.24) we can infer that
for $\epsilon<<1$,
$$\eqalignno{
  \rho &= 1 + {\cal O}(\epsilon^2)  \ , &(A.1a)\cr
  \psi &= {\cal O}(\epsilon^2) \ .  &(A.1b)\cr
}
$$
For fixed $t$ and small $r$, $\epsilon \sim r$ so (A.1) gives
$$\eqalignno{
  \rho &= 1 + {\cal O}(r^2) \ ,&(A.2a)\cr
  \psi &= {\cal O}(r^2) \ ,  &(A.2b)\cr
}
$$
while for fixed $r$ and large $|t|$, $\epsilon \sim t^{-2}$ so
(A.1) gives
$$\eqalignno{
  \rho &= 1 + {\cal O}(t^{-4}) \ ,&(A.3a)
  \cr\noalign{\vskip 0.2cm}
  \psi &= {\cal O}(t^{-4}) \ . &(A.3b)\cr
}
$$
Note that (A.2) is more stringent than just
the finite energy bounds (2.19).

Working near $\wfin$ when $r$ is large, the finite energy condition
places only weak restrictions on $\rho$ and $\psi$. However, to
solve (2.24) near $\wfin=\pi/2$ we must have $\rho\to\pm1,0$.
Writing $w=\wfin - \epsilon$, the asymptotic form for small
$\epsilon$ in the case that $\rho\to\pm 1$ becomes, as in (A.1),
\smallskip
$$\eqalignno{
  \rho &=  \pm 1 +
  {\cal O}(\epsilon^2)\ , &(A.4a)\cr
  \psi &=  {\cal O}
  (\epsilon^2)\ , &(A.4b)\cr
}
$$
whereas if $\rho\to 0$ we get
$$\eqalignno{
  \rho &=  a ~\epsilon^\eta \ ,
                             &(A.5a) \cr
  \psi &= \psi_0 + b ~\epsilon^\xi\ ,&(A.5b)
  }
$$
where $\eta$ and $\xi$ are as yet undetermined. We examine
the $\rho\to 0$ case.
Substituting (A.5) in (2.24b) gives $\xi = 2\eta$.
After substituting  (A.5) into  (2.24a)
we find two complex solutions for $\eta$ when $b\ne 0$,
$$
  \eta_\pm = {1\over 2}\bigg[1\pm i\sqrt{3+16\psi_0^2}~\bigg] \ .
\eqno(A.6)
$$
Complex solutions imply oscillatory behavior; however,
since the equations are non-linear (for small $\rho$ the
$1/\rho^3$ in (2.24a) is important) we cannot superimpose
these solutions to form real ones.
It turns out that we can linearize (2.24a) in $\rho$
in the following manner. To leading order in $\epsilon$,
(2.24b) implies that
$$
  {\psi '(\epsilon) \over \rho^2 (\epsilon)}=-{2 \psi_0  \over \epsilon}
  \ ,\eqno(A.7)
$$
where a prime denotes differentiation with respect to  $\epsilon$.
Using this form for $\psi '$ in (2.24a) we have to leading order
in $\epsilon$
$$
  \rho'' \ + \ {1 + 4 \psi_0^2 \over \epsilon^2}\ \rho =0
\ .\eqno(A.8)
$$
The general solution to (A.8) is
$\rho = a_{+} \epsilon^{\eta_{+}} + a_{-} \epsilon^{\eta_{-}}$,
where $a_{\pm}$ are arbitrary constants and
$\eta_{\pm}$ are given by (A.6).
We can now form
a real solution by appropriate linear combinations. We write
$$
  \rho = A \epsilon^{1/2} \cos[c~ {\rm ln}\epsilon + B] \ ,
\eqno(A.9)
$$
where $c=\sqrt{3+16\psi_0}/2$ and $A,B$ are arbitrary
constants. Substituting (A.9) back into (A.7)
and integrating gives
$$
  \psi = \psi_0 - A^2 \psi_0 ~\epsilon~\bigg[1+{\cos(2c~
  {\rm ln}\epsilon + 2B)\over 1+4c^2 } + {2 c ~\sin(2c~{\rm
ln}\epsilon+2B)\over 1 + 4c^2}\bigg] \ .
\eqno(A.10)
$$
The solutions found numerically in Section II appear to possess
this behavior. Only two of the constants $A$,
                                             $B$ and $\psi_0$ in
(A.10) are independent due to the two parameter nature of finite
energy solutions.

We now prove the assertion that for finite energy solutions of
(2.24), if $\psi$ is non-zero then it is monotonic, that is to
say $\psi$ is non-increasing or non-decreasing.
We prove this by contradiction. From the finite energy boundary
condition (A.1b), $\psi(w)$ must vanish at the  point $\winit=-\pi/2$.
Let $w^*$ be the first point greater than $\winit$
for which $\psi$ ceases to be monotonic. This means  $\psi$ is
either non-increasing or non-decreasing from $\winit$ to
$w^*$. Assume the latter (which is consistent with
Fig.~4); the former case may be handled in
a similar manner. Then $w^*$ is a relative maximum of $\psi(w)$,
{\it i.e.} $\psi'(w^*)=0$ and $\psi(w^*)> \psi(w)$ for
points $w$ slightly to the left and right of $w^*$.
Since $\psi$ is non-zero with vanishing derivative at $w^*$,
(2.24) implies that $\rho(w^*)=0$. We write $w=w^*+\epsilon$ and
expand $\rho$ and $\psi$ in powers of $\epsilon$:
$$\eqalignno{
  \rho &= a_1\epsilon + a_2\epsilon^2+a_3\epsilon^3+\ldots \ \ ,
  &(A.11a) \cr
  \psi &= b_0 + b_2\epsilon^2 + b_3\epsilon^3 +
  b_4\epsilon^4+\ldots \ \ . &(A.11b)
  }
$$
For $\psi$ to have a relative maximum at $w^*$, the leading
non-trivial $\epsilon$ dependence  must be an even power of
$\epsilon$ with a {\it negative} coefficient. We will now show that
this can not happen.

Assume that $a_1\ne 0$. We will relax this assumption in a moment.
Note that $b_0>0$ since $\psi$ started out zero and never decreases
until $w^*$. Using (A.11) in (2.24b) we find

$$\eqalignno{
  b_2 &= 0 \ \ , &(A.12a)\cr
  b_4 &= {3 a_2 b_3 \over 2a_1} + {b_0 a_1^2 \over 2\cos^2w^*} \ \ .
  &(A.12b) \cr
 }
$$
while $b_3$ is left undetermined.
We are not interested in the case in which $b_3 \ne 0$, since
the leading $\epsilon$ behavior would then be an odd power in
$\epsilon$. So if $b_3=0$, then (A.12b) gives $b_4>0$ since $b_0>0$.
This is a contradiction since the leading epsilon dependence must
be an even power of $\epsilon$ with {\it negative} coefficient.
If $a_1$ had vanished, let $a_n$ be the first non-zero coefficient
of (A.12a). Then similar reasoning would give
$\psi = b_0+|b_{2n+2}|\epsilon^{2n+2} + \ldots$, which again is a
contradiction. Therefore, $\psi(w)$  is monotonic over the whole
$w$-range.

A corollary of this result is that the only finite energy
solutions that interpolate between $\rho=+1$ and $\rho=-1$
as $w$ varies from $\winit$ to $\wfin$ have $\psi$ equal
to zero.
This can be seen as follows. Finite energy solutions start
at $\rho=1$ and $\psi=0$. From (A.4), if $\rho$ does not
asymptote to zero as $w\to\wfin$ then $\psi$ must return
to zero. This can not happen for non-vanishing $\psi$ which
never decrease.
\vfill
\eject
\centerline{\bf REFERENCES}
\medskip
\item{1.}  S. Adler, {\it Phys. Rev.} {\bf 177} (1969) 2426;
  J. Bell and R. Jackiw, {\it Nuovo Cimento} {\bf 51} (1969) 47;
  W.A. Bardeen, {\it Phys. Rev.} {\bf 184} (1969) 1848.
\medskip
\item{2.}  G. 't Hooft,{\it Phys. Rev.} {\bf D14} (1976) 3432; (E) {\bf D18}
  (1978) 2199.
\medskip
\item{3.}  A. Belavin, A. Polyakov, A. Schwarz and  Yu. Tyupkin,
  {\it Phys. Lett.} {\bf B59} (1975) 85.
\medskip
\item{4.}  N. Manton, {\it Phys. Rev.} {\bf D28} (1983) 2019;
  F. Klinkhamer and  N. Manton, {\it Phys. Rev.} {\bf D30} (1984) 2212.
\medskip
\item{5.}V. Kuzmin, V. Rubakov and
M. Shaposhnikov, {\it Phys. Lett.} {\bf B155}
  (1985) 36.
\medskip
\item{6.}  A. Ringwald, {\it Nucl. Phys.} {\bf B330} (1990) 1.
\medskip
\item{7.}  M. Mattis, {\it Phys. Rep.} {\bf 214} (1992) 159;
  V.V. Khoze, in {\it QCD at 200 TeV}, eds. L. Cifarelli and Yu.
  Dokshitzer (Plenum Press 1992).
\medskip
\item{8.}N. Christ, {\it Phys. Rev.} {\bf D21} (1980) 1591.
\medskip
\item{9.}  E. Farhi, V.V. Khoze and R. Singleton, MIT-CTP-2139 (1992).
\medskip
\item{10.}  M. L\"uscher, {\it Phys. Lett.} {\bf B70}   (1977) 321.
\medskip
\item{11.}  B. Schechter, {\it Phys. Rev.} {\bf D16} (1977) 3015.
\medskip
\item{12.}  R. Jackiw and C. Rebbi, {\it Phys. Rev. Lett.} {\bf 37} (1976)
  172; C. Callan, R. Dashen and D. Gross, {\it Phys. Lett.} {\bf B63}
  (1976) 334.
\medskip
\item{13.}  E. Witten,    {\it Phys. Rev. Lett.} {\bf 38} (1977) 121.
\medskip
\item{14.}  B. Ratra and L.G. Yaffe, {\it Phys. Lett.} {\bf B205}  (1988) 57.
\medskip
\item{15.}
  V. de Alfaro, S. Fubini and G. Furlan, {\it Phys. Lett.} {\bf B65}
  (1976) 163; J. Cervero, L. Jacobs and C. Nohl, {\it Phys. Lett.}
  {\bf B69} (1977) 351.
\medskip
\item{16.}  S. Coleman, {\it Commun. math. Phys.} {\bf 55} (1977) 113;
  S. Coleman and L. Smarr, {\it Commun. math. Phys.} {\bf 56}
  (1977) 1.
\vfill
\eject
\centerline{\bf FIGURE CAPTIONS}
\medskip
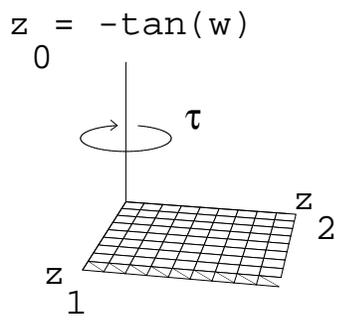
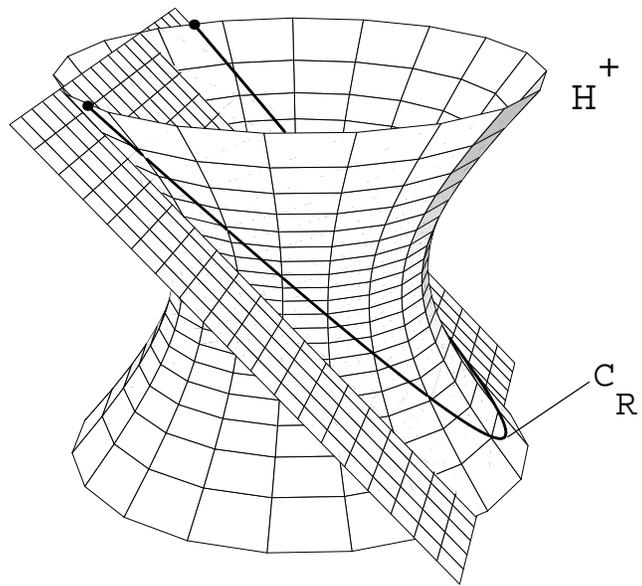
\item{Fig.~1}The hyperboloid $z_0^2-z_1^2-z_2^2=-1$.  The $r$,$t$
  coordinates cover the half of the hyperboloid, $H^+$, above the
  $z_0+z_2=0$ plane. We also show a typical fixed-$r$ contour
  $C_R$.
\medskip
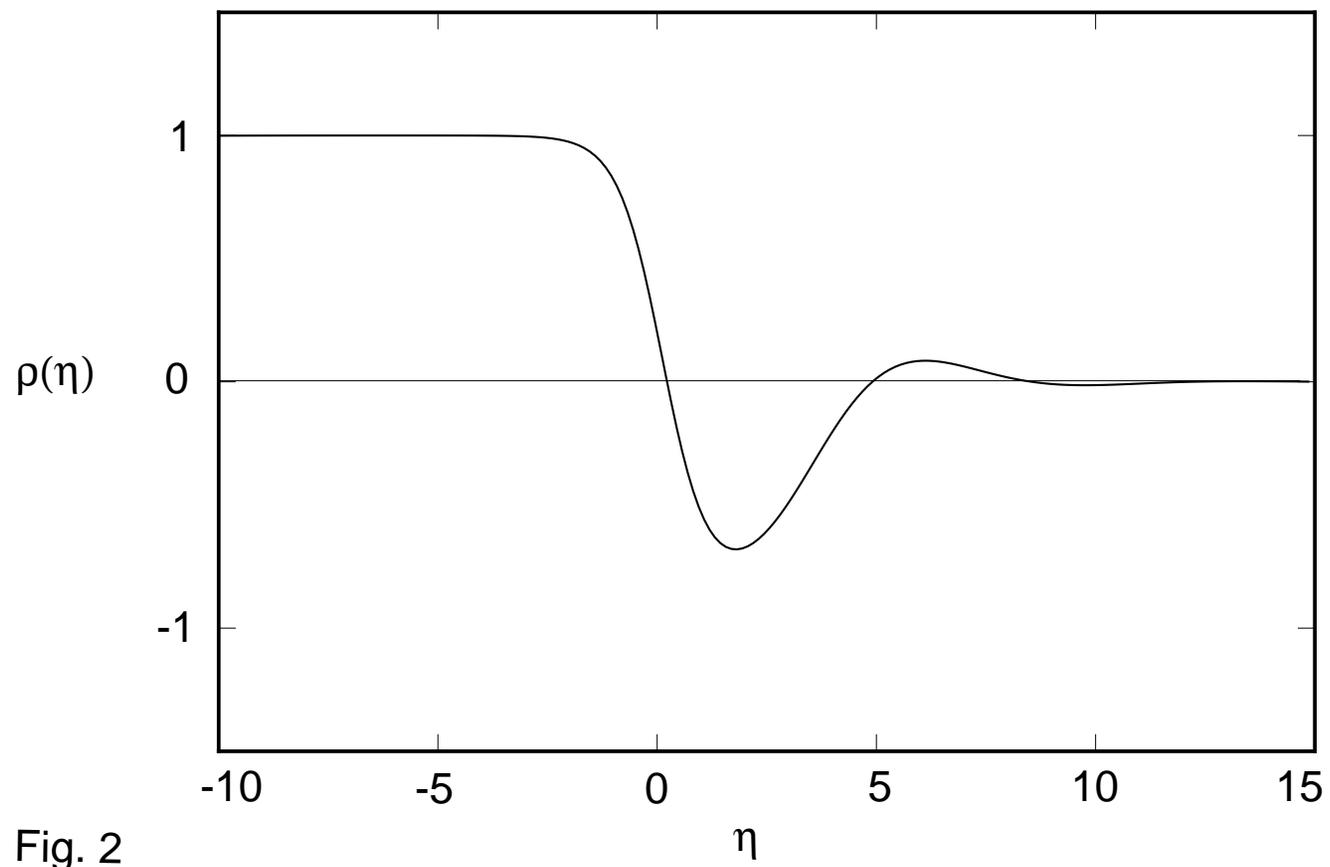
\item{Fig.~2:}The function
  $\rho(\eta)$ for the initial conditions $\rho(0)=0.2$
  and $\dot\rho(0)=-0.881146$. For a given $\rho(0)$, the velocity
  $\dot\rho(0)$ must be fine-tuned to ensure the finite energy
  condition $\rho\to1$ as $\eta\to-\infty$.
\medskip
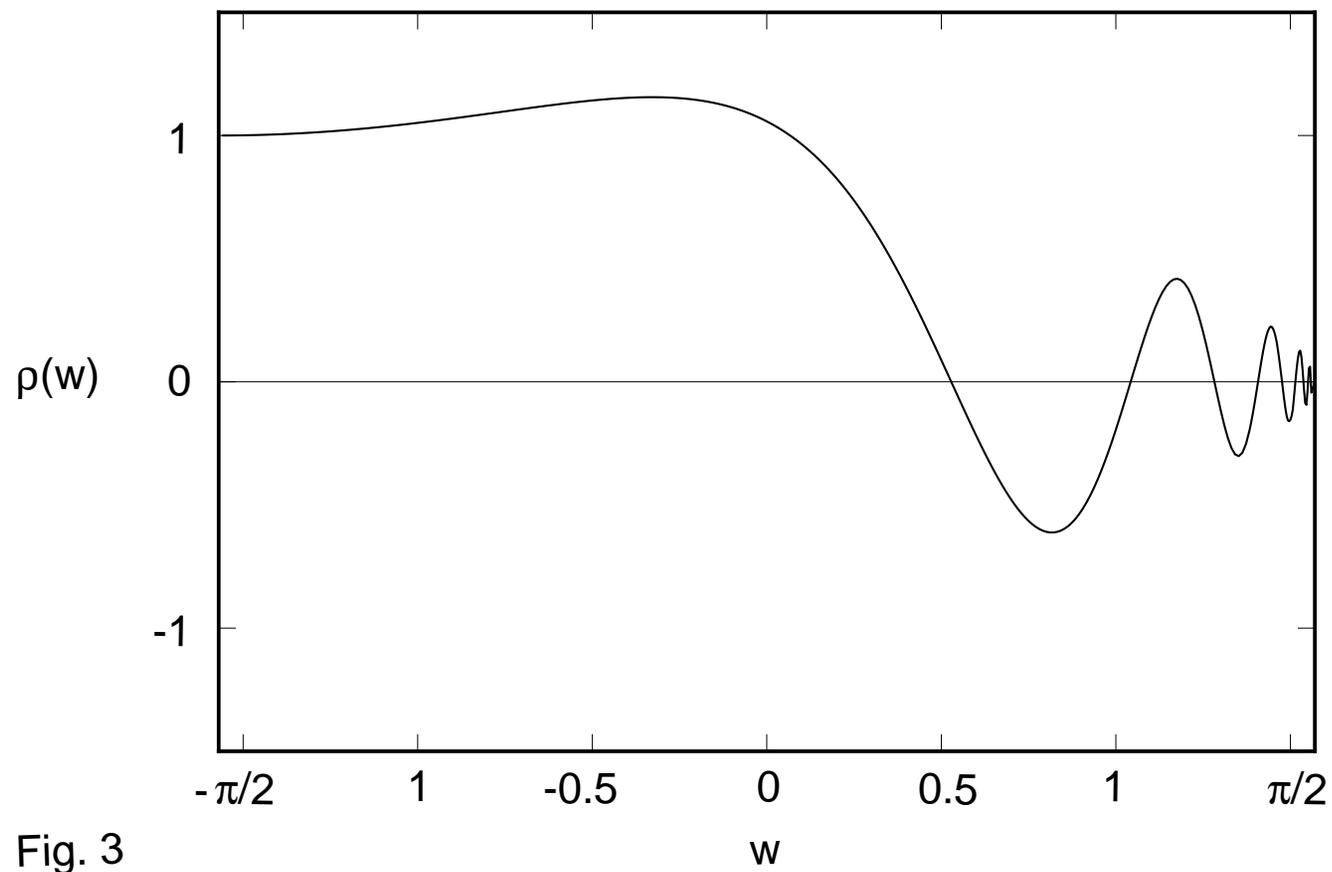
\item{Fig.~3:}A typical $\rho(w)$ for the solutions of Section II.
  We have chosen the initial conditions $\rho(w_{\rm i})=0.99995$,
  $\rho'(w_{\rm i})=0.01$, $\psi(w_{\rm i})=-0.0001$ and
  $\psi'(w_{\rm i})=0.02$. To avoid the singularity in (2.24)
  at $-\pi/2$ we have taken $w_{\rm i}=-\pi/2 + 10^{-2}$. In all
  graphs related to Section II, we will use the initial conditions
  stated here.
\medskip
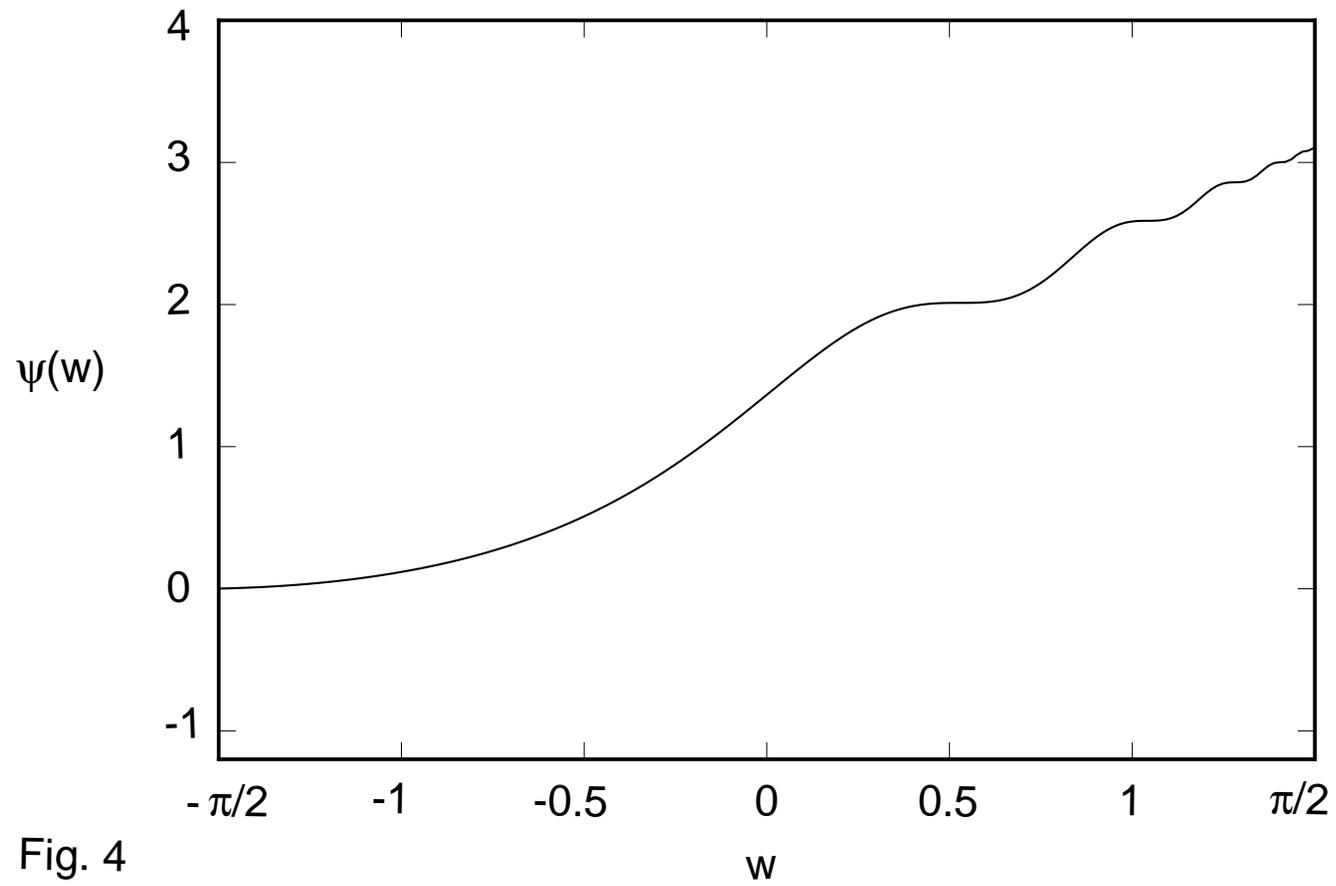
\item{Fig.~4:}A typical $\psi(w)$ for the solutions of Section II.
\medskip
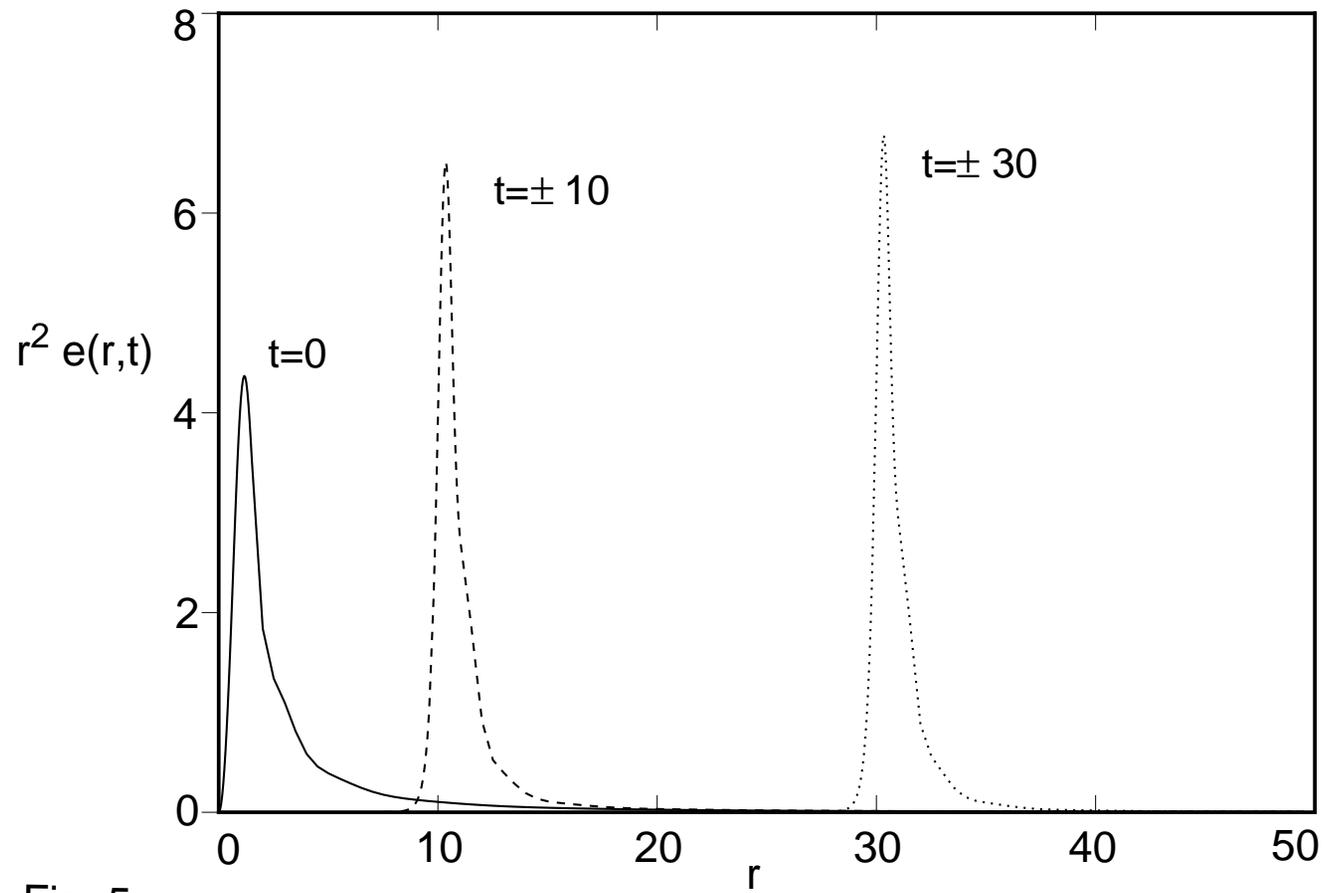
\item{Fig.~5:}Profiles of the energy density, $e(r,t)$,
  times $r^2$ in units of $8\pi/g^2$ for a sequence of times
  for a typical solution of Section II.
\medskip
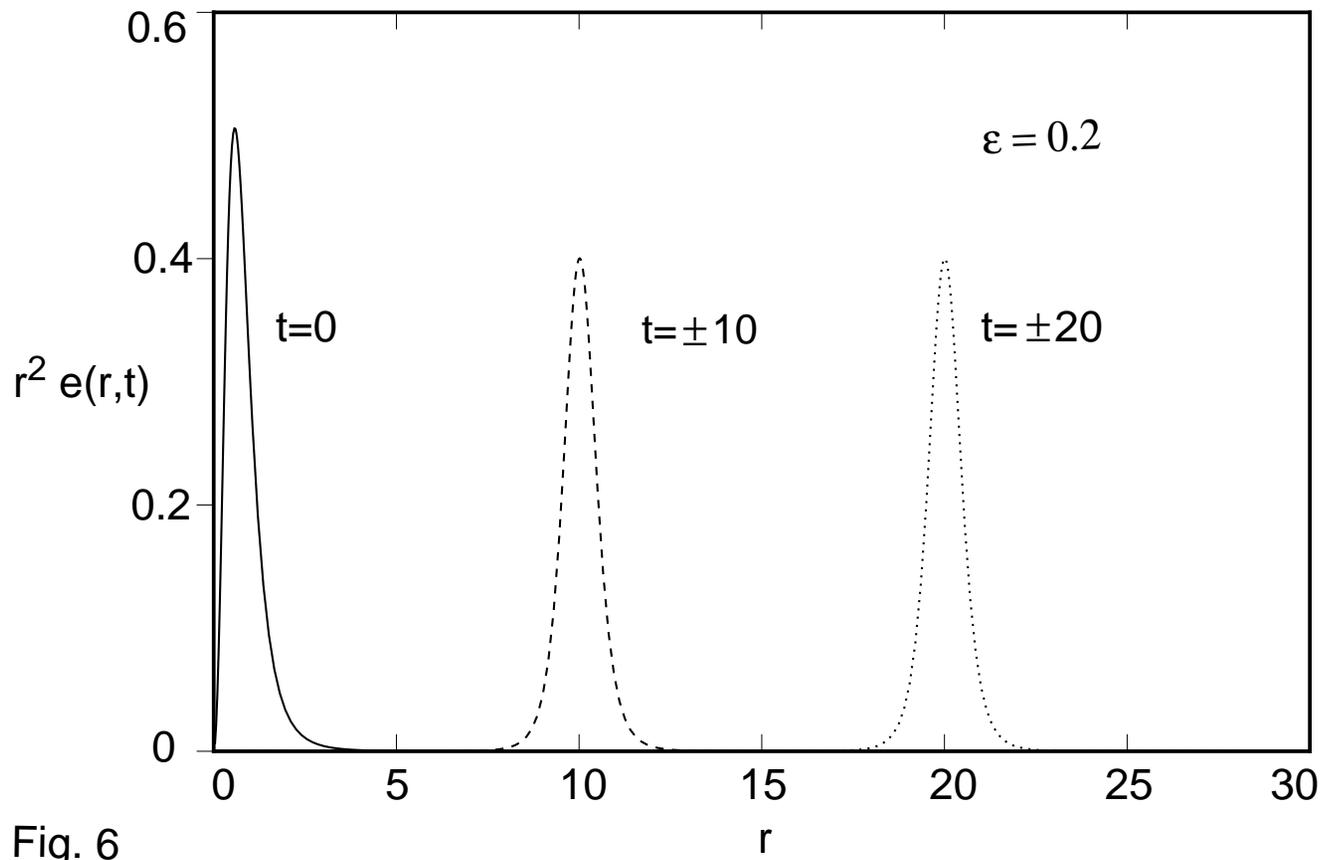
\item{Fig.~6.}Profiles of the energy density, $e(r,t)$, times
  $r^2$ in units of $8\pi/g^2$ for for a sequence times for a
   L\"uscher-Schechter solution with $\varepsilon=0.2$.
\medskip
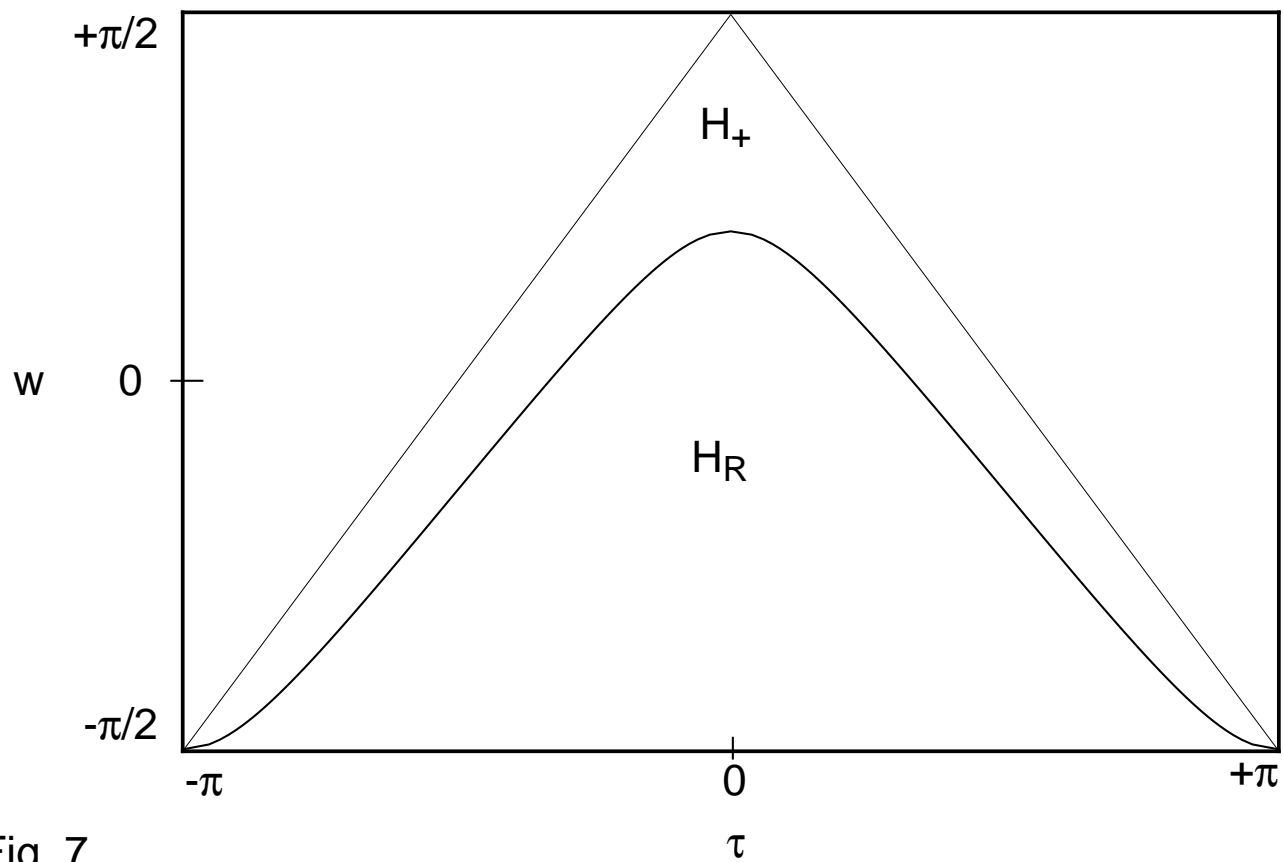
\item{Fig.~7.}The regions $H^+$ and $H_R$ in the $w$-$\tau$ plane.
\medskip
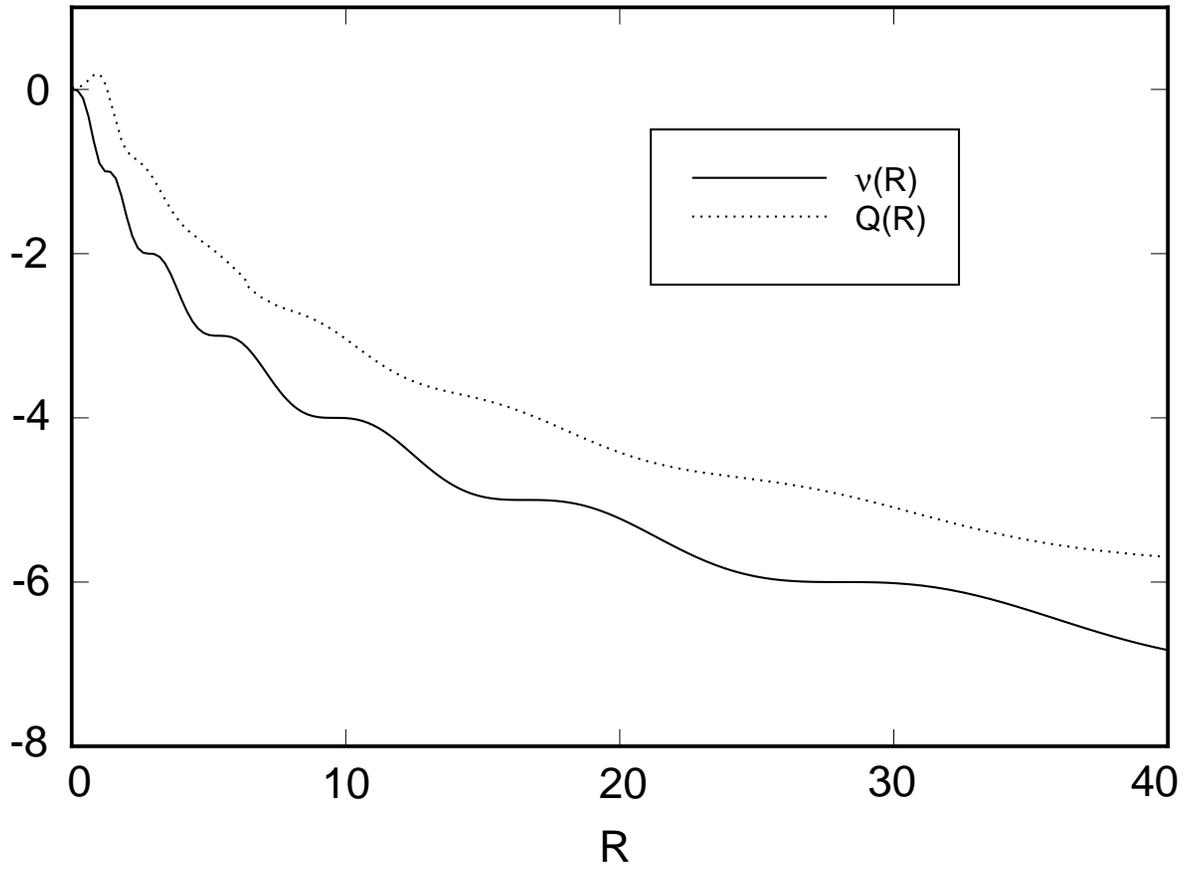
\item{Fig.~8:}The local winding number $\nu(R)$ and
  the local topological charge $Q(R)$ vs. $R$ for a typical solution
  of Section II.
\medskip
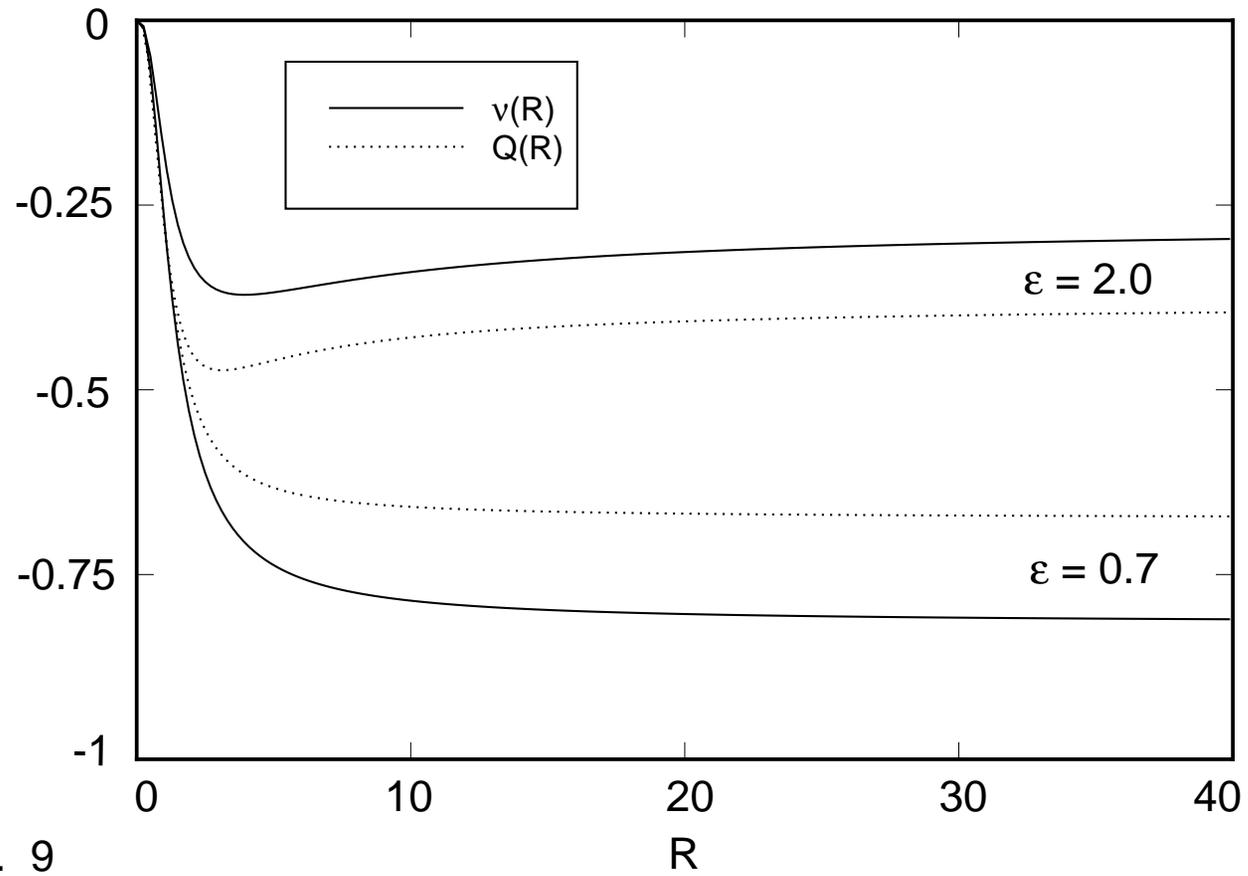
\item{Fig.~9:}The local winding number $\nu(R)$ and the local topological
  charge $Q(R)$ vs. $R$ for the L\"uscher-Schechter solutions with two
  values of $\varepsilon$ and $\tau_0=1$.
\medskip
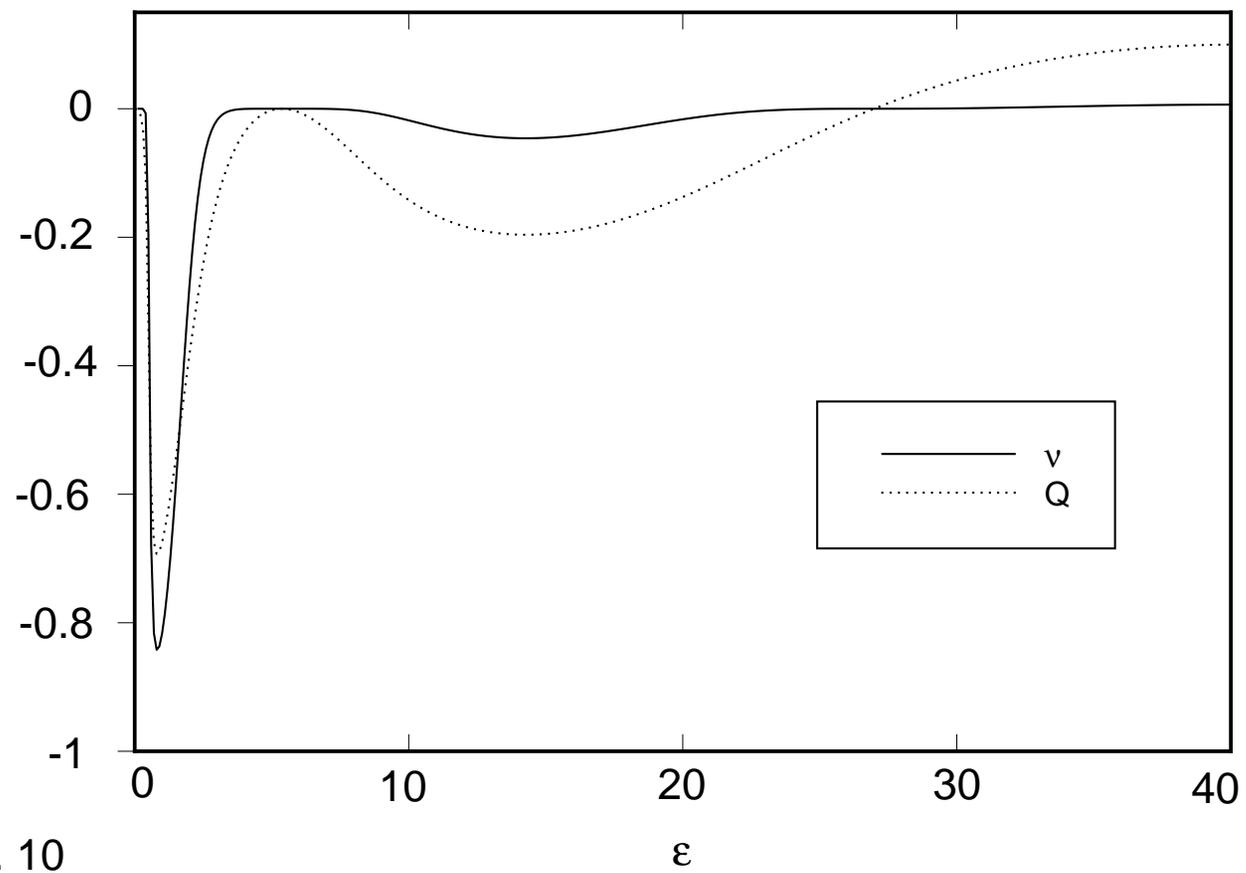
\item{Fig.~10:}The asymptotic winding number $\nu$ and the
topological charge, $Q$, for the L\"uscher-Schechter solutions
as a function of $\varepsilon$ with $\tau_0=1$.
\medskip
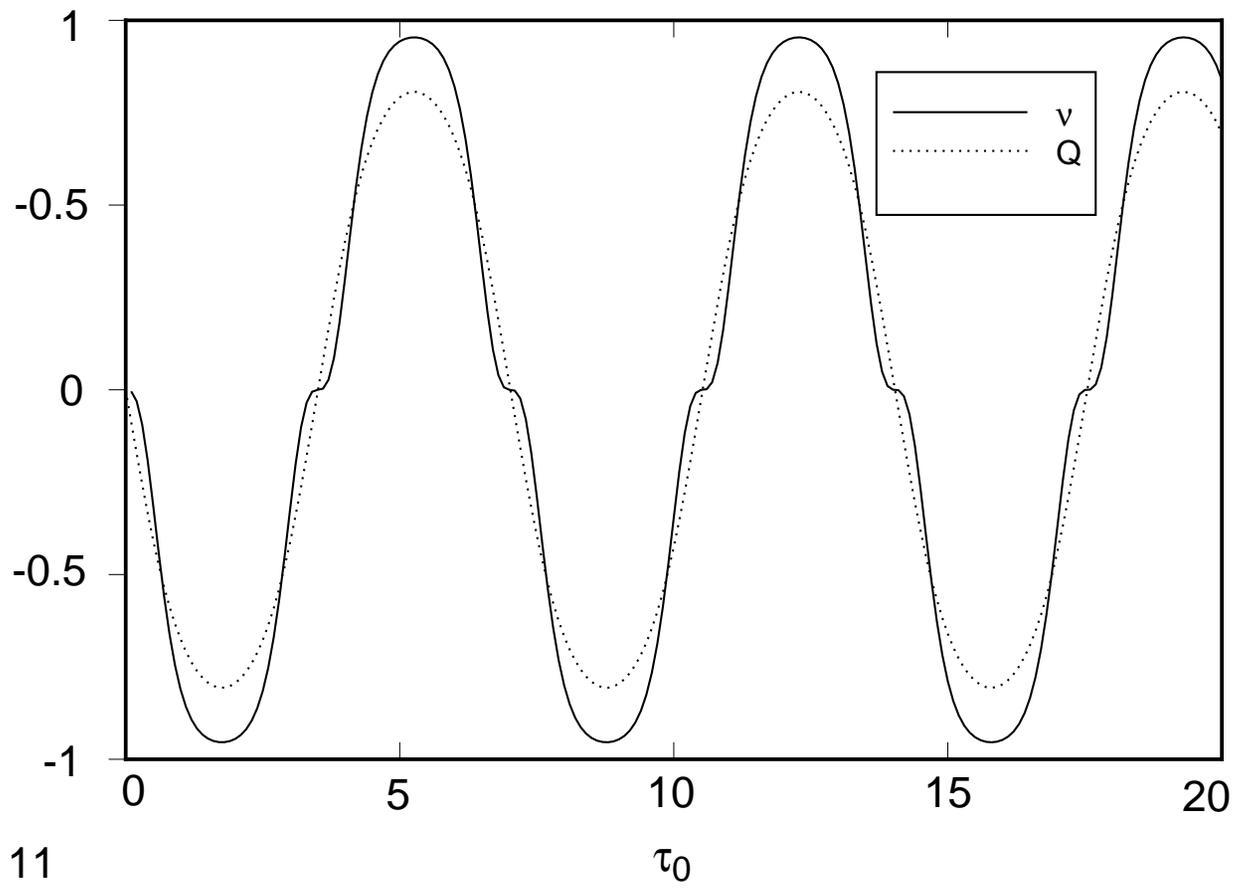
\item{Fig.~11:}The asymptotic winding number $\nu$ and the
  topological charge, $Q$, for the L\"uscher-Schechter solutions
  as a function of $\tau_0$ with $\varepsilon=0.7$.
\par
\vfill
\end